# Development of a Generalizable Data-driven Turbulence Model: Conditioned Field Inversion and Symbolic Regression

Chenyu Wu[1], Shaoguang Zhang[1], Yufei Zhang[1,*]

(1. School of Aerospace Engineering, Tsinghua University, Beijing, 100084, China)

**Abstract**

This paper addresses the issue of predicting separated flows with Reynolds-averaged Navier–Stokes (RANS) turbulence models, which are essential for many engineering tasks. Traditional RANS models usually struggle with this task, so recent efforts have focused on data-driven methods such as field inversion and machine learning (FIML) to correct this issue by adjusting the baseline equations. However, these FIML methods often reduce accuracy in attached boundary layers. To address this issue, we developed a "conditioned field inversion" technique. This method adjusts the corrective factor $\beta$ (used by FIML) in the shear–stress transport (SST) model. It multiplies $\beta$ with a shield function $f_d$ that is off in the boundary layer and on elsewhere. This maintains the accuracy of the baseline model for the attached flows. We applied both conditioned and classic field inversion to the NASA hump and a curved backward-facing step (CBFS), creating two datasets. These datasets were used to train two models: SR-CND (from our new method) and SR-CLS (from the traditional method). The SR-CND model matches the SR-CLS model in predicting separated flows in various scenarios, such as periodic hills, the NLR7301 airfoil, the 3D SAE car model, and the 3D Ahmed body, and outperforms the baseline SST model in the cases presented in the paper. Importantly, the SR-CND model maintains accuracy in the attached boundary layers, whereas the SR-CLS model does not. Therefore, the proposed method improves separated flow predictions while maintaining the accuracy of the original model for attached flows, offering a better way to create data-driven turbulence models.

## 1. Introduction

Reynolds-averaged Navier–Stokes (RANS) equations are widely used in daily engineering applications. The equations are used to compute the mean velocity and pressure and, in the case of compressible flow, the density. These equations rely on RANS turbulence models to describe flow turbulence. The RANS method is computationally more efficient than high-fidelity methods such as direct numerical simulation (DNS), large eddy simulation (LES), and detached eddy simulation (DES). This makes the RANS method especially valuable in aerodynamic optimization and other computationally intensive computational fluid dynamics (CFD) tasks. However, the RANS method has persistent shortcomings in accurately predicting separated flows, primarily due to the various assumptions and simplifications inherent in its modeling process. Consequently, improving the ability of RANS models to predict separated flows is crucial, as this would significantly expand their applicability and effectiveness in a broader range of engineering problems.

---

[*] Corresponding author, e-mail: zhangyufei@tsinghua.edu.cn

Recently, the field of RANS turbulence modeling has embraced data-driven methods [1][2]. Among these approaches, machine learning (ML) models have emerged as a booster for traditional CFD [3][4], particularly in addressing the complex challenge of separated flows faced by RANS turbulence models [5]~[12]. A notable example of such an approach is the field inversion and machine learning (FIML) method introduced by Duraisamy et al. [13]. The FIML approach involves adding a spatially varying correction factor, referred to as $\beta(\mathbf{x})$, to the standard turbulence model equations. This factor serves as a measure of the model's error. For instance, this correction factor might be applied as a multiplier to the destruction term in the shear–stress–transport (SST) model's $\omega$ equation. The next step involves solving an optimization problem, where the objective is to minimize the discrepancy between the RANS model's predicted quantity of interest (QoI) and the QoI derived from either high-fidelity methods or experimental data. This data assimilation process is also known as field inversion. Once this optimization is complete, a machine learning model is developed using the optimized $\beta(\mathbf{x})$ values. This model is designed to link local flow characteristics to the correction factor. The resultant ML model can then be applied in new CFD simulations to refine the turbulence model predictions. Variations of the FIML method have been documented in other studies [14][15]. A key advantage of the FIML data assimilation step is its ability to infer the entire $\beta(\mathbf{x})$ field using only sparse high-fidelity data [16], such as experimental measurements, thus ensuring the wide applicability of the FIML method. The data assimilation step in FIML also ensures that the subsequent ML step does not necessarily have to learn high-fidelity Reynolds stress directly, bypassing many numerical issues caused by ML-predicted Reynolds stress [17][18]. FIML has been used with considerable success in separated flow modeling tasks. Singh et al. [19] built a model using a multilayer perceptron (MLP) model with FIML and predicted the stall of wind turbine airfoils. Yan et al. [20][21] also used FIML with an MLP to address iced airfoils and 3-D separation flows around NASA's FAITH Hill project [22]. O. Bidar et al. [23] used a novel eigenspace perturbation method to place sensors in the region of highest uncertainty during field inversion, which increased the efficiency of data assimilation. Yan et al. [24] incorporated turbulence fluctuation features with FIML through a generative model, achieving performance gains in the prediction of flow separation in periodic hills. Andrea et al. [27] modified the corrective-multiplier's expression, making it more suitable for the augmentation of turbomachinery flows.

While FIML models have shown success in modeling flows similar to those in their training sets, challenges in their ability to generalize remain a concern. The generalizability of FIML models can be classified into four distinct levels, as detailed in Table 1.

Table 1. L1 to L4 generalizability of the FIML model

|    | *Definition* | *Example* |
|----|---|---|
| **L1** | The model performs well in a series of geometries similar to the training set. | The model is trained on one periodic hill and can be generalized to other periodic hills with different aspect ratios, such as [20]. |
| **L2** | The model does not negatively affect the baseline model's accuracy in simple wall-attached flows. | The model is trained on some separated flow cases and can be as accurate as its baseline model on the zero-pressure-gradient flat plate, such as [32] |
| **L3** | The model performs well in test cases that have separation features similar to the training set but completely different geometries and Reynolds numbers. | (1) The model is trained on a case where separation is caused by a blunt geometry and is tested to be effective in other cases that have completely different blunt geometries and Reynolds numbers, such as [29].<br>(2) The model is trained on a case where separation starts from a smooth surface (caused by a negative pressure gradient) and is tested to be effective on other smooth surface separations that have completely different geometries and Reynolds numbers. |
| **L4** | The model performs well in a series of test cases with completely different separation features, geometries, and | The model can accurately predict separation caused by blunt geometry as well as separation starting from a smooth surface. |

| Reynolds numbers. |
|---|

Current research on FIML, as referenced in previous studies [19][20][21][23][24][27], primarily focuses on testing Level 1 (L1) generalizability. Some models have demonstrated promising results in achieving Level 3 (L3) generalizability [28][29]. However, L2 generalizability, which pertains to a model's ability to maintain accuracy in simpler wall-attached flows, is critically important. The significance of L2 generalizability cannot be overstated, especially given that real-world scenarios often involve both simple wall-attached flows and separated flows, which substantially influence key metrics such as the drag coefficient. Consequently, Spalart suggested that fulfilling L2 generalizability should be treated as a hard constraint for data-driven turbulence models [30]. However, a notable gap in L2 generalizability was highlighted by Rumsey et al. [31], who found that a FIML model trained on an adverse-pressure-gradient (APG) flat plate did not perform accurately on a zero-pressure-gradient (ZPG) flat plate. This suggests that FIML models developed using traditional approaches might lack effectiveness in L2 scenarios. To address this issue, Joel Ho et al. [32] introduced a probabilistic FIML model. This model deactivates the corrective factor in situations of high uncertainty, as determined by Gaussian process regression-based predictions. This approach showed improved performance on the ZPG flat plate and NACA0012 airfoil without flow separation (indicating enhanced L2 generalizability), although it did compromise the model's effectiveness in improving separated flows (i.e., L1 and L3 generalizability). Another strategy to boost L2 generalizability was pursued by Jäckel [33], who incorporated a strong inductive bias into the FIML model's structure. Jäckel implemented an interpretable, closed-form Gaussian radial basis function (RBF) to maintain the accuracy of the baseline Spalart Allmaras (SA) model in ZPG flat plate scenarios. However, this approach, possibly due to the strong inductive bias of the RBF, did not sufficiently reduce the baseline SA model's error in separated flows, leading to reduced effectiveness in terms of L1 and L3 generalizability. Recently, a novel method called constrained recalibration proposed by Bin et al. [34][35] has increased the data-driven turbulence model's L2 generalizability nicely. The physical constraints related to the law of the wall are extracted as algebraic relations and are preserved during the data-augmentation stage for the separated flows. Bin et al. [34][35] show that the law of wall can indeed be preserved no matter how the constrained recalibration model is adjusted, showing strong L2 generalizability. By now, the adjustable parameters in the constrained recalibration model remain constant throughout the field. The FIML method can be incorporated in the future to gain a better L3 generalizability.

This paper introduces a new approach named conditioned field inversion (FI-CND) aimed at enhancing the L2 generalizability of FIML models while maintaining L1 and L3 generalizability. This method builds upon the classic field inversion (FI-CLS) technique, retaining its basic algorithm but innovating in how the corrective factor $\beta(\mathbf{x})$ is applied. Our focus is on refining the SST model. The FI-CLS method directly multiplies the corrective factor $\beta(\mathbf{x})$ with the destruction term of the $\omega$ equation. In contrast, the proposed FI-CND method takes a different approach. The proposed FI-CND method multiplies the destruction term with $[\beta(\mathbf{x}) - 1]f_d + 1$, where $f_d$ is a shielding function introduced by Spalart et al. [36] for robust detached eddy simulation (DES). This function, $f_d$, is zero within the boundary layer and one outside it. This means that the corrective factor $\beta(\mathbf{x})$ is only active outside the boundary layer and is filtered out inside the boundary layer, thereby maintaining the accuracy of the baseline model for wall-attached flows (L2 generalizability). Essentially, this approach assumes that the baseline model accurately calculates the attached boundary layer and requires no modification in this region. The filter-based approach has been used in data-driven turbulence modeling literature to form a zonal turbulence model with non-linear Reynolds stress correction [25], to detect the region with an adverse pressure gradient [26], or to reduce the computational cost [27].To the best of the authors' knowledge, there's no work in the current literature that uses the filter-based approach in the field inversion stage of FIML and uses it to preserve the accuracy of the wall-attached flows. Both FI-CLS and FI-CND were applied to the NASA hump case and the curved-backward-facing step (CBFS) case. These applications produced two separate datasets for $\beta(\mathbf{x})$, each combining data from both cases. Following these steps, two closed-form models for $\beta$ were developed using symbolic regression: SR-CLS for the FI-CLS dataset and SR-CND for the FI-CND dataset. Symbolic regression was chosen in the machine-learning phase of FIML due to its strong performance in separated flows or L3 generalizability [28][29]. The results showed that both the SR-CLS and SR-CND models significantly improved the performance in separated flows, including on different configurations such as periodic hills, the NLR-7301 two-element airfoil, the 3D notchback SAE model, and the 3D Ahmed body. These results demonstrate that the FI-CND method does not reduce the L3 generalizability of models based on FI-CLS. Moreover, only the SR-CND model accurately predicted the skin friction coefficient in the zero-pressure-gradient (ZPG) flat plate and the NACA0012 airfoil at a zero angle of attack. This outcome highlights the effectiveness of our FI-CND method in achieving L2 generalizability, a critical aspect for more accurate and reliable turbulence modeling in engineering applications.

## 2. Frameworks of conditioned field inversion and symbolic regression

In this section, we introduce the FIML framework that is applied to generate data-driven turbulence models. We focus on augmenting the SST model. Figure 1 provides a visual guide for our methodology. We first perform field inversion, including the classic version and the conditioned version proposed in this paper, to obtain the corrective factor $\beta(\mathbf{x})$ from sparse high-fidelity data, forming a dataset of $\beta$ and its corresponding local flow features denoted as vector $\mathbf{w}$. Then, we use symbolic regression to distill a mapping from $\mathbf{w}$ to $\beta$ based on the dataset output by field inversion. Finally, the expression $\beta(\mathbf{w})$ is integrated back into the turbulence model, replacing $\beta(\mathbf{x})$, making the resultant model ready to be applied to new cases. The following subsections provide a more detailed description of the field inversion step and the symbolic regression step.

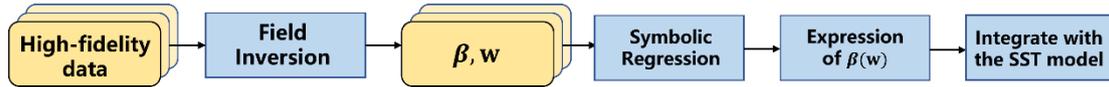

Figure 1. Basic flowchart of the FIML framework

## 2.1. Classic field inversion and conditioned field inversion

The baseline RANS model we aim to modify is the SST 2003 model [37] (abbreviated as SST hereafter), which is widely used for engineering purposes. The transport equations of the SST model are as follows:

$$\frac{\partial(\rho k)}{\partial t} + \frac{\partial(\rho u_j k)}{\partial x_j} = P - \beta^* \rho \omega k + \frac{\partial}{\partial x_j}\left[(\mu + \sigma_k \mu_t)\frac{\partial k}{\partial x_j}\right]$$
$$\frac{\partial(\rho \omega)}{\partial t} + \frac{\partial(\rho u_j \omega)}{\partial x_j} = \frac{\gamma}{\nu_t} P - \theta \rho \omega^2 + \frac{\partial}{\partial x_j}\left[(\mu + \sigma_\omega \mu_T)\frac{\partial \omega}{\partial x_j}\right] + 2(1 - F_1)\frac{\rho \sigma_{\omega 2}}{\omega}\frac{\partial k}{\partial x_j}\frac{\partial \omega}{\partial x_j} \quad (1)$$

where $k$ is the turbulent kinetic energy, $\omega$ is the specific dissipation rate, and $\mu_t$ is the eddy viscosity. For a more detailed description of the SST model, the readers are referred to [37].

In classic field inversion (FI-CLS), such as [29], the spatially varying corrective factor $\beta(\mathbf{x})$ ($\mathbf{x}$ is the spatial coordinate) is directly multiplied by the destruction term of $\omega$'s equation:

$$\frac{\partial(\rho\omega)}{\partial t} + \frac{\partial(\rho u_j \omega)}{\partial x_j} = \frac{\gamma}{\nu_t}P - \beta(\mathbf{x})\theta\rho\omega^2 + \frac{\partial}{\partial x_j}\left[(\mu + \sigma_\omega \mu_T)\frac{\partial \omega}{\partial x_j}\right] + 2(1-F_1)\frac{\rho \sigma_{\omega 2}}{\omega}\frac{\partial k}{\partial x_j}\frac{\partial \omega}{\partial x_j} \quad (2)$$

Eq. (2) allows $\beta(\mathbf{x})$ to be modified anywhere in the flow field, including the attached boundary layer. It is reduced to the baseline model if $\beta(\mathbf{x}) = 1$ everywhere. However, varying $\beta(\mathbf{x})$ from its baseline value 1, particularly in the attached boundary layer, can disrupt the calibration of the original model's constants, such as $\gamma$ and $\theta$, which are crucial for accurate turbulence modeling [34][35]. For example, in the original model (Eq. (1)), $\gamma$, $\theta$, and $\sigma_\omega$ are calibrated to fulfill the following necessary condition for accurate resolution of the log layer [38]:

$$\theta - \beta^*\gamma = \kappa^2\sqrt{\beta^*}\sigma_\omega, \kappa = 0.41 \quad (3)$$

If we want the classic field inversion model in Eq. (2) to calculate the log layer accurately, a similar necessary condition must hold:

$$\beta(\mathbf{x})\theta - \beta^*\gamma = \kappa^2\sqrt{\beta^*}\sigma_\omega, \kappa = 0.41 \quad (4)$$

By comparing Eqs. (3) and (4), it becomes clear that $\beta(\mathbf{x})$ must be equal to 1 within the logarithmic layer for the conditions of Eq. (4) to be satisfied. Therefore, any deviation of $\beta(\mathbf{x})$ from 1 in the logarithmic layer would compromise the model's ability to accurately represent the boundary layer calibration. In addition, other calibrations, such as the wake-law calibration for boundary layers with different pressure gradients [38], are also negatively affected by $\beta(\mathbf{x}) \neq 1$ in the boundary layer. The following structure is proposed for the multiplier of $\omega$'s destruction term in an attempt to solve the problem shown above:

$$\frac{\partial(\rho\omega)}{\partial t} + \frac{\partial(\rho u_j \omega)}{\partial x_j} = \frac{\gamma}{\nu_t}P - [(\beta(\mathbf{x}) - 1)f_d + 1]\theta\rho\omega^2 + \frac{\partial}{\partial x_j}\left[(\mu + \sigma_\omega \mu_T)\frac{\partial \omega}{\partial x_j}\right]$$
$$+ 2(1-F_1)\frac{\rho \sigma_{\omega 2}}{\omega}\frac{\partial k}{\partial x_j}\frac{\partial \omega}{\partial x_j} \quad (5)$$

Similarly, Eq. (5) is reduced to the baseline model's $\omega$ equation when $\beta(\mathbf{x}) = 1$ everywhere. The shielding function $f_d$ was constructed by Spalart et al. in [36] and has the following form:

$$f_d = 1 - \tanh[(8r_d)^3], \quad r_d = \frac{\mu + \mu_T}{\rho\kappa^2 d^2 \sqrt{u_{i,j}u_{i,j}}} \quad (6)$$

where $d$ is the wall distance and $u_{i,j} = \partial u_i/\partial x_j$. $f_d$ is constructed such that it is 0 inside the boundary layer and 1 outside. In practical use, the attached boundary layer typically falls within the region where the shielding function $f_d$ is zero, as illustrated in Figure 2. Implementing $f_d$ in Eq. (5) ensures that the destruction term remains $\theta\rho\omega^2$ in this shielded area. This means that within the attached boundary layer, the model functions identically to the baseline SST model, thereby preserving all of its original calibrations. Outside this shielded region, the corrective factor $\beta(\mathbf{x})$ comes into play, modifying the destruction term to $\beta(\mathbf{x})\theta\rho\omega^2$. This approach is justified, as research [39][40] has indicated that inaccuracies in RANS turbulence models, especially in predicting flow separation, mostly arise in the separated shear layer—an area outside the shielded region. This method of applying $\beta(\mathbf{x})$ in Eq. (5) is called conditioned field inversion (FI-CND). It maintains the reliability of the baseline model in the boundary layer while allowing for adjustments where the model typically faces uncertainty.

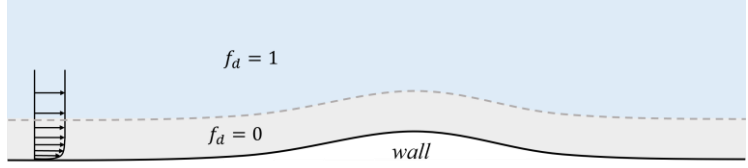

Figure 2. Typical $f_d$ distribution

For both FI-CLS and the FI-CND, the optimized $\beta(\mathbf{x})$ that can minimize the difference between the RANS prediction and high-fidelity data is determined by solving the following optimization problem (in discrete form):

$$\min_{\boldsymbol{\beta}} J = \lambda_{QoI} \sum_{i=1}^{K}[d_i - h_i(\boldsymbol{\beta})]^2 + \lambda_{L_2} \sum_{j=1}^{N}(\beta_j - 1)^2 \tag{7}$$

$\beta_j$ is $\beta(\mathbf{x})$'s value in the $j^{th}$ cell in the CFD simulation, and $\boldsymbol{\beta}$ is the vector whose $j^{th}$ element is $\beta_j$. $d_i$ is the $i^{th}$ high-fidelity quantity-of-interest (QoI) data point, such as the $x$-velocity obtained by LES or DNS. $h_i(\boldsymbol{\beta})$ is the RANS-predicted QoI data point induced by a specific $\beta(\mathbf{x})$ distribution represented discretely by $\boldsymbol{\beta}$. $K$ is the number of sparse high-fidelity samples, and $N$ is the number of cells used in the CFD simulation. The first term of $J$ measures the difference between the high-fidelity data and the RANS result, and the second term represents the deviation of $\beta(\mathbf{x})$ from its baseline value 1. $\lambda_{QoI}$ and $\lambda_{L_2}$ are two constants. In this study, $\lambda_{QoI} \approx \left[\sum(d_i - h_i(1))^2\right]^{-1}$ to make $J \approx 1$ when $\beta_j = 1$ in each cell. $\lambda_{L_2}$ is approximately $1 \times 10^{-5} \sim 1 \times 10^{-4}$. While ensuring a decent drop in the error of QoI, the value of $\lambda_{L_2}$ is made as large as possible to reduce the deviation of $\beta$ from its baseline value 1. Some trial and error might be needed to find a good value for $\lambda_{L_2}$. According to our experience, making the regularization term as large as 10%~20% of the error of QoI after optimization is often enough.

We use the sequential least squares programming (SLSQP) [41] method to solve the optimization problem in Eq. (7). Since the optimization problem involves functions defined by partial differential equations (the RANS-predicted QoI data point depends on $\beta$ through the solution of the RANS equations and the turbulence model equations), we use the discrete adjoint method [42] to compute $J$'s gradients to $\beta$. The outline of the discrete adjoint method is briefly described as follows. Suppose $R$ is the residual of the discretized governing equations (the RANS equations and the turbulence model equations). $R$ can be viewed as a vector-valued function having $m \cdot N$ outputs ($m$ is the number of governing equations, $N$ is the number of grid cells), and $R$'s inputs are $\beta$ and the flow variables (such as velocity, $k$, and $\omega$) in each cell $q$:

$$R(q, \beta) \tag{8}$$

Similarly, $J$ can also be viewed as a scalar function of $q$ and $\beta$:

$$J(q, \beta) \tag{9}$$

In a converged CFD simulation, the CFD solver maps a given $\beta$ to its corresponding $q$, so the converged solution of the CFD solver is a function of $\beta$: $q = q(\beta)$. The function is defined implicitly by the condition that $q$ and $\beta$ should fulfill the discretized governing equations:

$$R(q(\beta), \beta) = 0 \tag{10}$$

Consequently, the gradient of $J$ can be expressed as:

$$\nabla_\beta J(q(\beta), \beta) = \frac{\partial J}{\partial \beta} + \frac{\partial J}{\partial q} \nabla_\beta q \tag{11}$$

$\nabla_\beta q$ can be obtained by taking the gradient of Eq. (10):

$$\nabla_\beta R(q(\beta), \beta) = \frac{\partial R}{\partial \beta} + \frac{\partial R}{\partial q} \nabla_\beta q = 0 \Rightarrow \nabla_\beta q = -\left(\frac{\partial R}{\partial q}\right)^{-1} \frac{\partial R}{\partial \beta} \tag{12}$$

Eq. (11) can then be rewritten as follows:

$$\nabla_\beta J(q(\beta), \beta) = \frac{\partial J}{\partial \beta} + \frac{\partial J}{\partial q} \nabla_\beta q = \frac{\partial J}{\partial \beta} - \frac{\partial J}{\partial q} \left(\frac{\partial R}{\partial q}\right)^{-1} \frac{\partial R}{\partial \beta} \tag{13}$$

The adjoint vector $\psi$ is defined as the solution of the following linear equation:

$$\left(\frac{\partial R}{\partial q}\right)^T \psi = -\left(\frac{\partial J}{\partial q}\right)^T \tag{14}$$

Then, Eq. (13) is equivalent to:

$$\nabla_{\boldsymbol{\beta}} J(\boldsymbol{q}(\boldsymbol{\beta}), \boldsymbol{\beta}) = \frac{\partial J}{\partial \boldsymbol{\beta}} + \boldsymbol{\psi}^T \frac{\partial \boldsymbol{R}}{\partial \boldsymbol{\beta}} \tag{15}$$

Based on the equations above, the discrete adjoint method applied to compute the gradient of $J$ to $\boldsymbol{\beta}$ can be summarized as follows:

1. After a converged CFD simulation, we compute the Jacobians $\partial \boldsymbol{R}/\partial \boldsymbol{q}$, $\partial J/\partial \boldsymbol{q}$, $\partial J/\partial \boldsymbol{\beta}$, and $\partial \boldsymbol{R}/\partial \boldsymbol{\beta}$.
2. We solve linear Eq. (14) to obtain the adjoint vector $\boldsymbol{\psi}$.
3. We compute $\nabla_{\boldsymbol{\beta}} J(\boldsymbol{q}(\boldsymbol{\beta}), \boldsymbol{\beta})$ using Eq. (15).

The Jacobians are calculated using automatic differentiation [43][44]. The entire framework for carrying out FI-CLS and FI-CND is built upon the open-source discrete adjoint solver DAFoam [45]~[47]. DAFoam uses OpenFOAM's [48] solver for forward evaluation (i.e., solving the CFD problem and obtaining the value of $J$) by applying the second-order finite volume method.

## 2.2. Symbolic regression

Field inversion provides us with only the numerical value of the optimized corrective factor $\beta$ field in a given case. We have to construct a function $\beta(\mathbf{w})$ that can map the local flow features $\mathbf{w}$ (such as the strain rate) to $\beta$ during CFD simulation. In this study, we use the symbolic regression (SR) algorithm to build a closed-form analytical expression for $\beta(\mathbf{w})$ based on the dataset generated by field inversion. A simple closed-form expression is pursued instead of typical black-box models such as multilayer perceptron and random forest due to its good efficiency, and interpretability [29][49]. We use the open-source software PySR written by Cranmer [50] to perform symbolic regression.

Five local flow features are selected as the input features for symbolic regression. Their definitions and physical meanings are listed in Table 2. The element functions that constitute the final expression are listed in Table 3. The basic operators like $+, -, \times, /$ and the functions that often appear in traditional turbulence models such as $\tanh(\cdot)$ are selected as element functions. A better set of element functions might be found but that is beyond the scope of this work. Every expression produced and evaluated by PySR consists of some element functions in Table 3 and features in Table 2. We choose the weighted mean square error (WMSE) as the loss function to evaluate the

performance of a given expression $E$, and the label is $\beta - 1$:

$$WMSE(E) = \frac{1}{M}\sum_{i=1}^{M} \alpha_i[(\beta_i - 1) - \hat{y}_i]^2 \tag{16}$$

$M$ is the total number of samples, $\alpha_i$ is the weight for the $i^{th}$ sample, and $\hat{y}_i$ is the predicted value of the $i^{th}$ sample by expression $E$.

During the training process, PySR maintains several populations of expressions and uses the genetic algorithm described in [50] to evolve the population, generating new expressions for each population. When the training is complete, PySR outputs the expression with the smallest WMSE at each complexity level $C \leq C_{max}$. $C_{max}$ is set to 16 in this study, and $C$ is calculated by:

$$C(E) = N_{opr} + 2N_{var} + 4N_{const.} \tag{17}$$

$N_{opr}$, $N_{var}$, and $N_{const}$ are the number of operators, the number of variables, and the number of constants in a given expression $E$, respectively. The weights of the variable and the constant are increased to encourage PySR to generate simpler expressions while keeping the number of constants and variables as small as possible.

Table 2. The local flow features chosen to construct the expression for $\beta(\mathbf{w})$.

| Name | Definition | Physical meanings |
|---|---|---|
| $\lambda_1$ | $\text{tr}(\widehat{\mathbf{S}}^2)$ | The 1st, 2nd, and 5th scaler invariances of nondimensional strain rate $\widehat{\mathbf{S}} = \mathbf{S}/(\beta^*\omega)$ and nondimensional rotation rate $\widehat{\mathbf{\Omega}} = \mathbf{\Omega}/(\beta^*\omega)$ derived by Pope [51] for the general tensor representation of the Reynolds stress. $\beta^*$ is a model constant of the SST model and equals 0.09. The 3rd and the 4th invariances are omitted because they are zero in 2D flows and our training set is 2D. Intuitively, $\lambda_1$, $\lambda_2$, and $\lambda_5$ measure the ratio of the mean strain/rotation frequency and the turbulence dissipation frequency ($\omega$). |
| $\lambda_2$ | $\text{tr}(\widehat{\mathbf{\Omega}}^2)$ | |
| $\lambda_5$ | $\text{tr}(\widehat{\mathbf{\Omega}}^2 \cdot \widehat{\mathbf{S}}^2)$ | |
| $\mathbf{Re}_\Omega$ | $|\Omega|d^2/\nu$ | A feature proposed by [52] and [16] to identify strong shear away from the wall. |
| $P_k/\epsilon$ | $\tau_{ij}^R u_{i,j}/(\beta^* k\omega)$ | The ratio between the production and the dissipation of turbulent kinetic energy. $\tau_{i,j}^R$ is the Reynolds stress tensor calculated using the Boussinesq hypothesis. [39][52] proposed empirically that the traditional turbulence models underestimate eddy viscosity when $P_k/\epsilon$ is large. |
| $\eta$ | $\lambda_2\lambda_5/\text{Re}_\Omega$ | The feature extracted by symbolic regression from the field inversion dataset of the CBFS case in a previous work [29]. The physical interpretation of this term is a bit complicated and can be found in [29]. |

Table 3. The list of element functions

| Operator Type | Operators |
|---|---|
| Unary | $\exp(\cdot), \tanh(\cdot), \dfrac{1}{1+(\cdot)}, \dfrac{1}{(\cdot)}$ |
| Binary | $(\cdot)+(\cdot), (\cdot)-(\cdot), (\cdot)\times(\cdot), \dfrac{(\cdot)}{(\cdot)}, \min(\cdot,\cdot), \max(\cdot,\cdot), (\cdot)^{(\cdot)}$ |

# 3. Field inversion and model training on the NASA hump case and the CBFS case

In this section, we apply both the classic field inversion (FI-CLS) and the conditioned field inversion (FI-CND) methods to the NASA hump [53], which features a relatively high Reynolds number ($\sim 10^6$), and the curved-backward-facing step (CBFS) [54], which features a low Reynolds number ($\sim 10^4$), using large eddy simulation (LES) data from [53][54] as our high-fidelity reference. We develop two unique datasets by applying the FI-CLS and FI-CND methods to data from both the NASA hump and CBFS cases. By combining data from these cases, we incorporate a range of physics from high to low Reynolds numbers, enriching the dataset with diverse aerodynamic behaviors. This diversity may boost the generalizability of models trained on these datasets. Subsequently, we employ symbolic regression on each dataset to create two predictive models: the SR-CLS model from the FI-CLS dataset and the SR-CND model from the FI-CND dataset, aiming for enhanced accuracy in aerodynamic predictions. This integration demonstrates the effectiveness of our training approach in improving model accuracy.

## 3.1. Field inversion

### 3.1.1. CBFS

The geometry and boundary conditions of the CBFS case are shown in Figure 3. The Reynolds number based on the step height $h = 1$ m is $\text{Re}_h = 13700$. The velocity at the center of the inlet is set to $1$ m/s. The objective function is shown in Eq. (18), with the LES $x$-velocity [54] being the high-fidelity data.

$$J = \lambda_{QoI} \sum_{i=1}^{30} [u_i^{LES} - u_i(\boldsymbol{\beta})]^2 + \lambda_{L_2} \sum_{j} (\beta_j - 1)^2 \tag{18}$$

The first summation is taken over the 30 randomly placed points in the blue-shaded region in Figure 3 (where the separation resides), denoted by blue triangles. $u_i^{LES}$ is the LES $x$-velocity at the $i^{th}$ extraction point, and $u_i(\boldsymbol{\beta})$ is its RANS counterpart. The second summation is taken over all the cells. The FI-CLS and the FI-CND share the same objective function (Eq. (18)). The initial design of the $\beta$ field is set to 1 uniformly. The convergence history of the optimization is shown in Figure 4. FI-CND decreases the objective function by 74.7%, while FI-CLS decreases it by 88.3%.

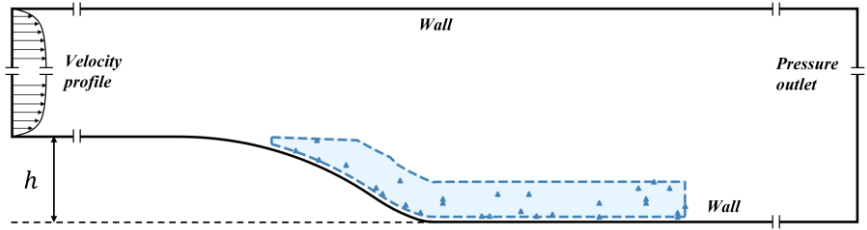

Figure 3. The geometry and boundary conditions of the CBFS case

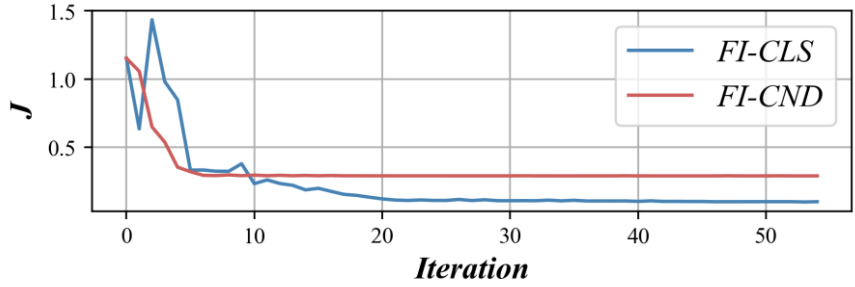

Figure 4. Convergence history of the objective function.

The streamline plot in Figure 5 demonstrates that both the FI-CLS and FI-CND approaches align the separation zone more accurately with the LES data than does the baseline SST model, which significantly overestimates the size of this zone. Furthermore, the velocity profiles in Figure 6 indicate that the $x$-velocity from both optimized $\beta$ fields is more consistent with the LES data, with FI-CLS showing a slightly closer match in accuracy.

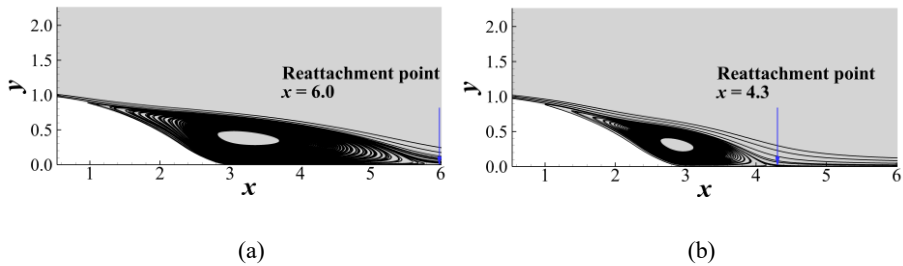

(a)          (b)

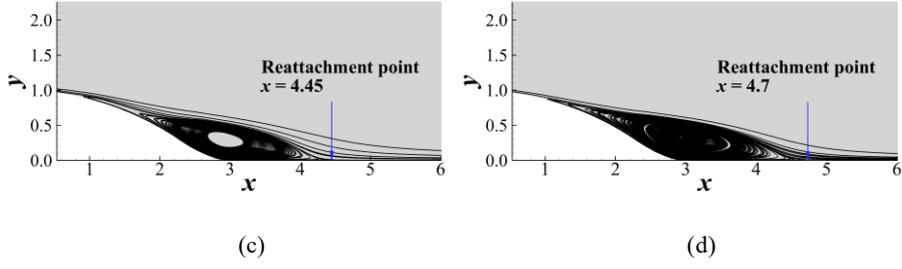

(c)                            (d)

Figure 5. Streamline plot of the separation zone. (a) SST; (b) LES; (c) FI-CLS; (d) FI-CND

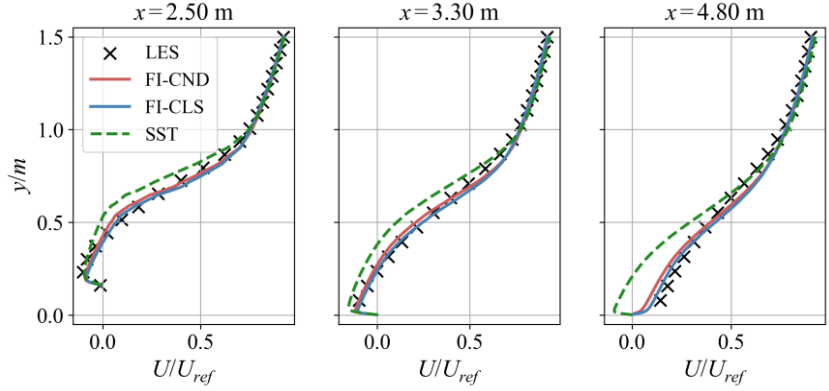

Figure 6. Velocity profiles at different $x$-standpoints.

The optimized $\beta$ fields generated by FI-CLS and FI-CND are shown in Figure 7. FI-CLS increases $\beta$ in both the separated shear layer and the attached boundary layer preceding the step, while FI-CND only increases $\beta$ in the separated shear layer. This demonstrates the ability of the conditioned field inversion method to protect the attached boundary layer.

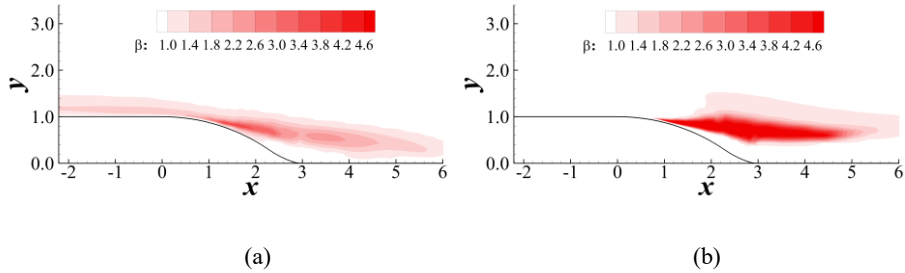

(a)                            (b)

Figure 7. $\beta$ distributions obtained by (a) FI-CLS and (b) FI-CND

### 3.1.2. NASA hump

The geometry and boundary conditions of the NASA hump case are shown in Figure 8. The Reynolds number based on the hump's chord length $c = 1 \text{ m}$ is $\text{Re}_c = 0.936 \times 10^6$. The inlet velocity is set to 14.88 m/s. The FI-CLS and the FI-CND share the same objective function shown in Eq.(18). The first summation is taken over 40 points marked by triangles in the blue-shaded region of Figure 8 to measure the difference between the $x$-velocity predicted by RANS and the LES data [53].

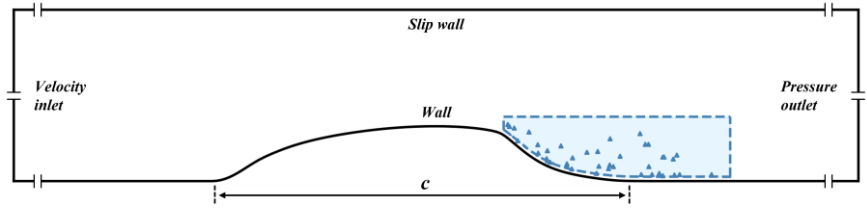

Figure 8. The geometry and boundary conditions of the NASA hump

We start with a uniform initial distribution of $\beta$ set to 1. Figure 9 illustrates the convergence history of the objective function for both FI-CLS and FI-CND. Both methods achieve convergence after approximately 140 SLSQP iterations. In terms of performance, the objective function shows a 66.7% reduction in FI-CND compared to a 61.7% reduction in FI-CLS. We observe a notable suppression of the separation zone for both FI-CND and FI-CLS compared to the baseline SST model after field inversion, aligning more closely with the LES data. This is demonstrated in Figure 10. Furthermore, a detailed comparison of the velocity field in Figure 11 reveals that the $x$-velocity from the optimized $\beta$ fields better matches the LES data, indicating an improvement in model accuracy.

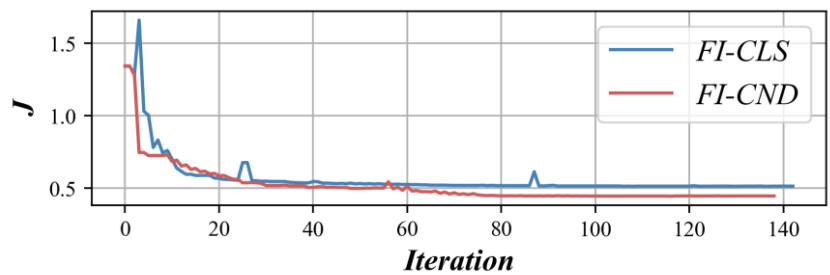

Figure 9. The convergence history of the field inversion process

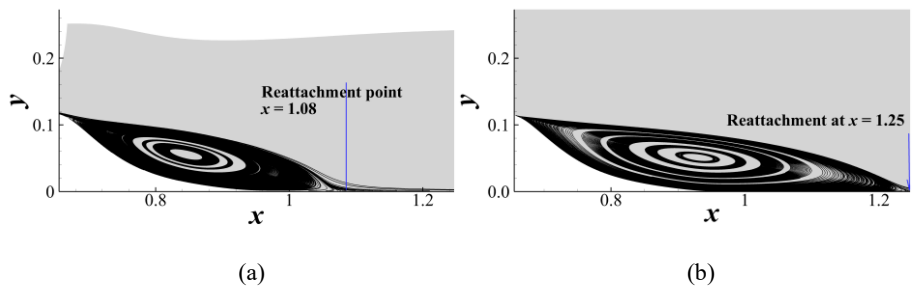

(a)                                                              (b)

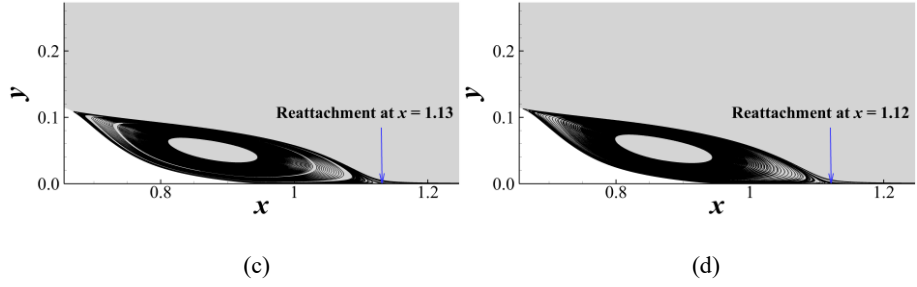

Figure 10. Streamline plot of the separation zone. (a) LES data, (b) baseline SST model, (c) FI-CLS, (d) FI-CND

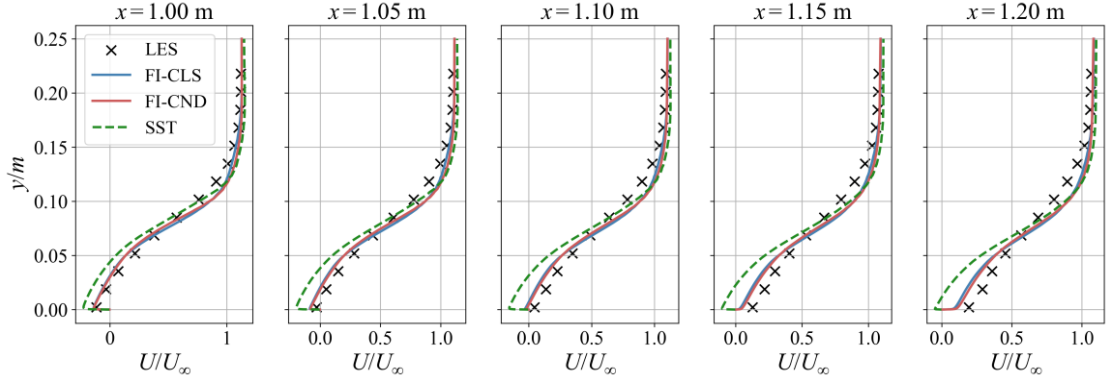

Figure 11. Velocity profiles at different downstream locations.

Figure 12 displays the optimized $\beta$ fields produced by both FI-CLS and FI-CND. In the FI-CLS approach, $\beta$ increases to approximately 2 near the separated shear layer. Notably, there is also an increase in $\beta$ within the attached boundary layer at the top of the hump. This increase in $\beta$ in the attached boundary layer could disrupt the calibration of the baseline SST model, leading to inaccuracies. On the other hand, the FI-CND method effectively safeguards the attached boundary layer, allowing $\beta$ to increase only around the separated shear layer, reaching values between 4 and 5. This outcome illustrates the efficacy of our conditioned field inversion technique in maintaining a $\beta$ value of 1 in the attached boundary layer, thereby preserving the integrity of the baseline SST model's calibration while addressing areas of high uncertainty.

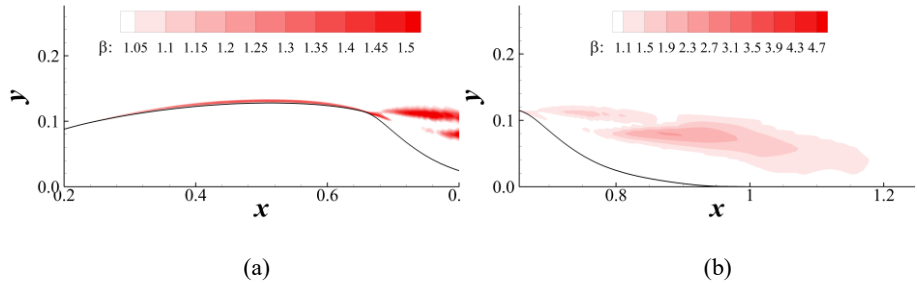

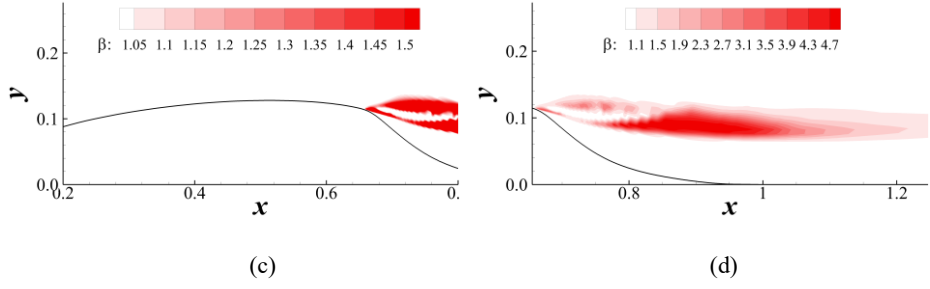

Figure 12. $\beta$ field obtained by field inversion. (a) FI-CLS, around the top of the hump; (b) FI-CLS, near the separated shear layer; (c) FI-CND, around the top of the hump; (d) FI-CND, near the separated shear layer

## 3.2. Symbolic model training

By mixing the field inversion results of the NASA hump case and the CBFS case, we obtain two datasets, including the local flow features described in Table 3 and the optimized corrective factor $\beta$. Note that a tuple of the local features and the corrective factor $(\mathbf{w}, \beta)$ is treated as one sample; therefore, the total number of samples equals the number of cells used in CFD analysis. One dataset is produced by FI-CLS, and the other is produced by FI-CND. We then downsample each mixed dataset to balance the proportion of trivial samples (where $|\beta - 1| < 0.05$) and nontrivial samples (where $|\beta - 1| \geq 0.05$), as shown in Figure 13. Each dataset contains approximately 3000 samples after downsampling. This downsampling is crucial for training efficiency, as larger datasets with approximately 10,000 samples can drastically slow the training with PySR, sometimes requiring dozens of hours. $\gamma$ in Figure 13 is the ratio between the trivial and the non-trivial samples in the original mixed dataset. That is, if we denote the number of non-trivial samples as $N$, then the number of trivial samples is:

$$N_{trivial} = \gamma N \qquad (19)$$

The weight in the loss function (Eq. (16)) is thus set based on the ratio $\gamma$:

$$\alpha_i = \begin{cases} 1, & i \in S_n \\ \gamma, & i \in S_t \end{cases} \qquad (20)$$

$S_n$ is the index set of the nontrivial samples, and $S_t$ is the index set of the trivial samples. The weight in Eq. (20) can offset the bias caused by the change in sample distribution after downsampling. Yan et al. [21] showed that the downsampling and weight adjustment procedure has little impact on the accuracy of the model built by the FIML framework.

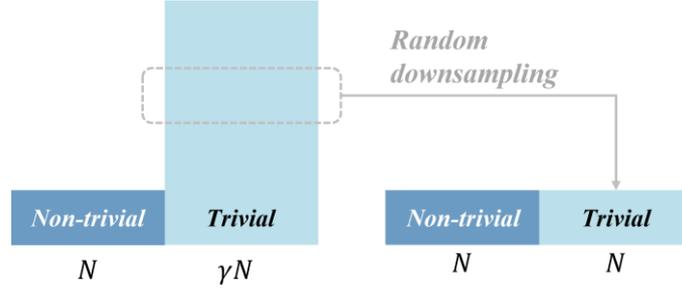

Figure 13. Dataset downsampling.

For the training on each dataset, the total number of populations is set to 20, and the number of expressions in each population is set to 80. The training lasts for 1000 iterations. After training on each dataset, we select the most accurate expression as our final choice. The chosen expressions are shown in Table 4. According to the definition of features in Table 2, the expressions suggest that $\beta$ will be increased where the non-dimensional rotation rate or strain rate is large (which often happens in the separated shear flows). An increase in $\beta$ would result in a decrease in $\omega$, and thus increasing eddy viscosity. The increased eddy viscosity will promote the momentum diffusion from the mainstream to the low-speed recirculation region, thus correcting the size of the separation. The physical implication of the correction term is consistent with the empirical findings in [39] and [40], which argue that the error of the traditional turbulence models mainly comes from the underestimated eddy viscosity in the separated shear flows far from the wall layer. Because the focus of the expression is outside of the wall layer (in the separated shear flow), it is still effective even with the filter function $f_d$ prevents the correction near the wall. The expression produced by FI-CLS is then integrated into Eq. (2), and the expression of FI-CND is integrated into Eq. (5). We call the two resultant models SR-CLS and SR-CND, respectively. Figure 14 shows the $\beta$ values predicted by SR-CLS and SR-CND using the features obtained by field inversion (without CFD iteration), indicating the effectiveness of symbolic regression in approximating the $\beta$ produced by field inversion. In this study, we found that the SR-CLS and the SR-CND model increase the total runtime (of the same CFD iteration steps) by about 15% and 10% respectively, which is acceptable. On the contrary, a study by Yin et al. [8] showed that if a random forest model is called in every iteration of CFD, the computational time required to converge the solution is about 30 times the convergence time of the baseline model. Consequently, the symbolic model is indeed more efficient.

Table 4. The final expressions trained on each dataset

| Training dataset | Expression |
|---|---|
| FI-CLS dataset | $\beta - 1 = -\dfrac{3}{500}\lambda_5 \tanh(-0.092\lambda_2)$ |
| FI-CND dataset | $\beta - 1 = \min(0.00435\lambda_2^2, 3.806)$ |

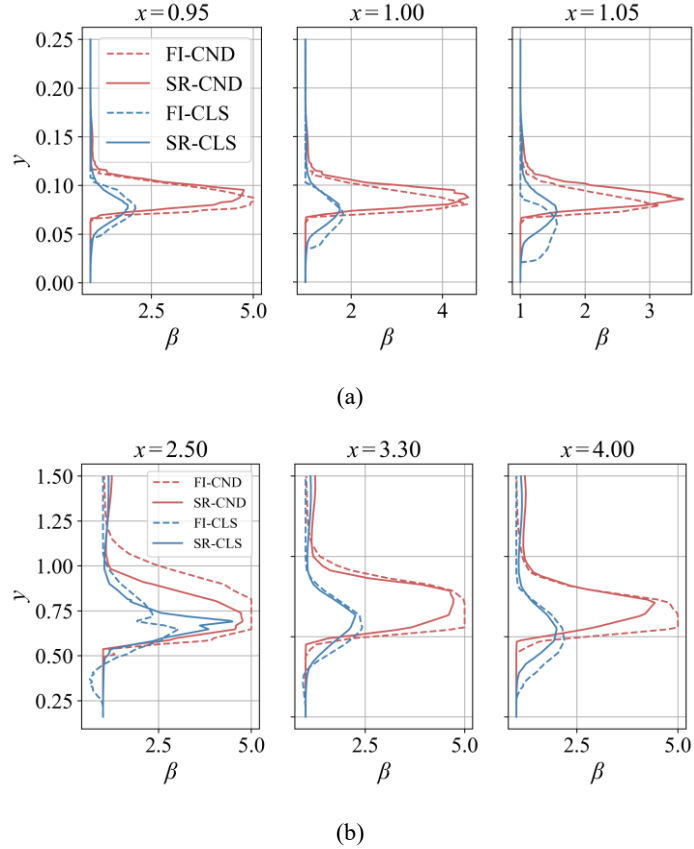

Figure 14. $\beta$ profiles predicted by the SR-CLS and SR-CND methods for (a) the NASA hump case and (b) the CBFS case

## 3.3. Application of the models on the training set

The SR-CLS and the SR-CND models are applied to the CBFS case and the NASA hump case to test the effectiveness of our training and the ability of the models to couple with CFD. The SR-CLS and the SR-CND models are called in every CFD iteration to update the $\beta$ field. Note that in this section and going forward, for the SR-CND model, we plot the distribution of $(\beta - 1)f_d + 1$ instead of $\beta$ since the flow field is directly influenced by the former. All calculations are performed using OpenFOAM's SimpleFOAM solver [48].

### 3.3.1. CBFS

Figure 15 features streamlines predicted by the SR-CLS and SR-CND models, showing an improvement over the reattachment point at $x = 6.0$, as given by the SST model. The velocity profiles in Figure 16 reveal that both the SR-CLS and SR-CND models align more closely with the LES data, although they slightly deviate from the field inversion results. This deviation could stem from the combined CBFS and NASA hump data used in training and the limited fitting capacity of the symbolic regression algorithm. The $C_f$ distribution, presented in Figure 17, indicates that the SR-CLS, SR-CND, and SST models yield similar results in the attached boundary layer and on the curved step. The key distinction is observed in the reattachment region, where the SST model tends to underpredict $C_f$. In contrast, both the SR-CLS and SR-CND models demonstrate improved accuracy in this area. Figure 18 shows that both the SR-CLS and SR-CND models exhibit increased $\beta$ values in the separated shear layer.

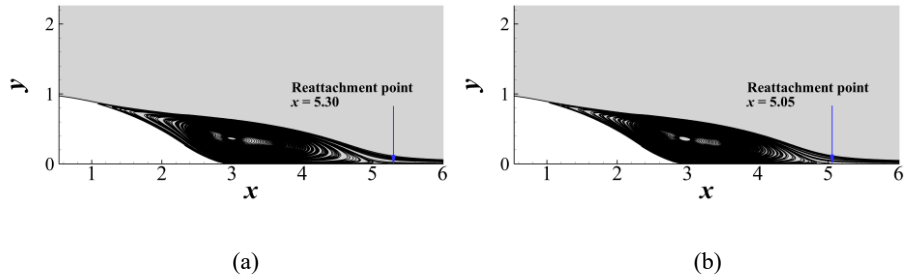

Figure 15. Streamline plot of the SR-CLS and the SR-CND models.

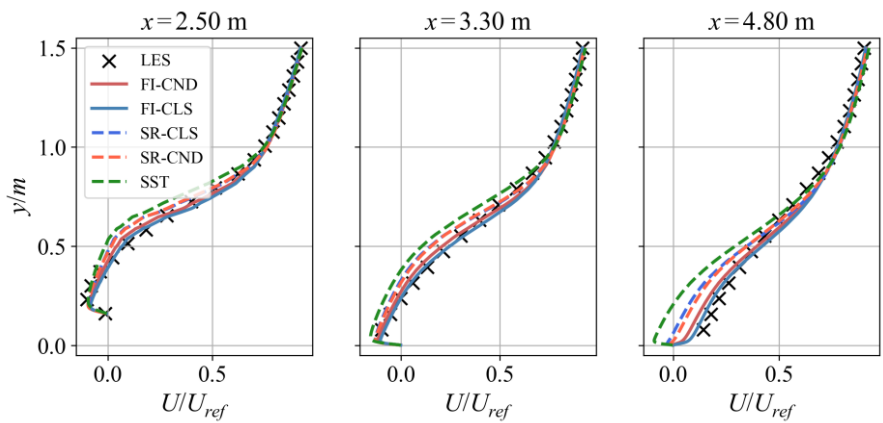

Figure 16. Velocity profiles at different $x$ standpoints

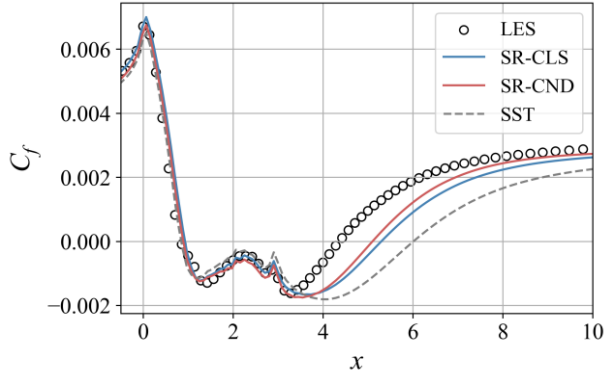

Figure 17. $C_f$ distribution on the curved step and downstream

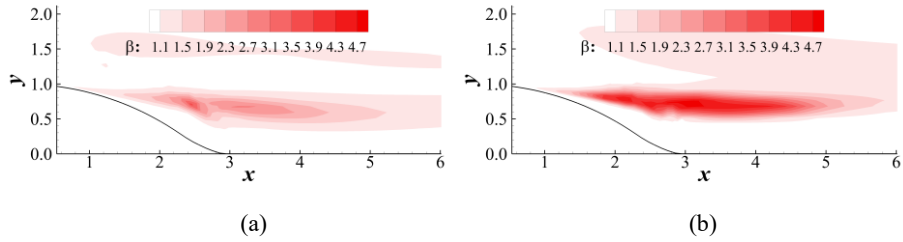

Figure 18. $\beta$ distributions given by (a) the SR-CLS model and (b) the SR-CND model

In summary, both the SR-CLS and SR-CND models outperform the baseline SST model, particularly in predicting the reattachment location, demonstrating the effectiveness of the training.

### 3.3.2. NASA hump

The streamline visualization in Figure 19 demonstrates that both the SR-CLS and SR-CND models more accurately predict the reattachment point in the separation zone than does the baseline SST model. The $x$-velocity profiles shown in Figure 20 further confirm the significant improvements achieved by both the SR-CLS and SR-CND models. The skin friction coefficient ($C_f$) distributions on and downstream of the hump ($x \in [0.0,1.0]$ and $(1.0, +\infty)$, respectively) are illustrated in Figure 21(a). Here, the SR-CLS and SR-CND closely match the experimental $C_f$ data [55] (measured at the same Reynolds number as the LES data used for field inversion) in the downstream separation zone, a region where the baseline SST model shows considerable deviation. However, while the predictions of the SR-CND model are similar to those of the SST model on top of the hump, the SR-CLS model slightly overestimates the $C_f$ value there. Figure 21(b) provides a detailed comparison of the $C_f$ error (relative to the experimental data) for each part. The mean squared errors (MSEs) of SR-CLS and SR-CND in the separation zone are approximately 6.7% of those of the SST model. At the top of the hump, the MSEs of the SR-CND and SST models are roughly equivalent, whereas the MSE of the SR-CLS model is 160% that of the SST model, indicating a greater deviation.

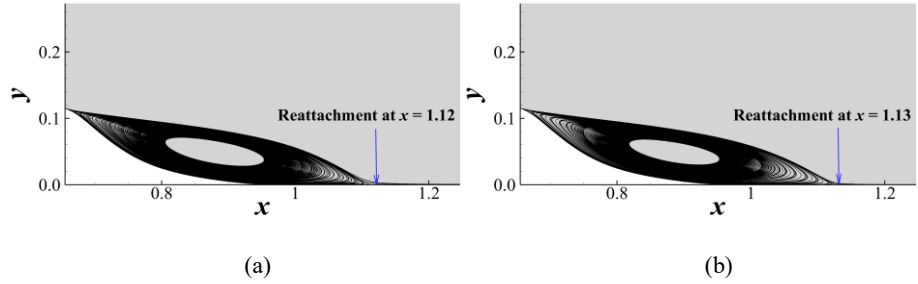

Figure 19. Streamlines in the separation zone given by (a) SR-CLS and (b) SR-CND

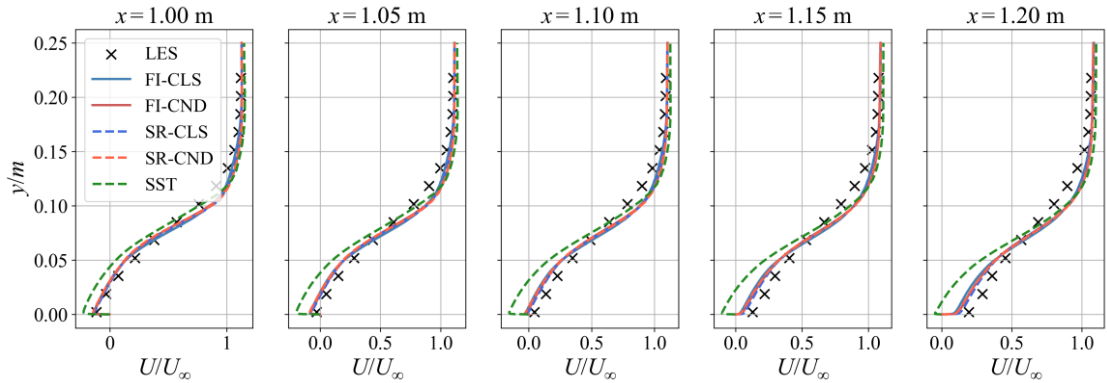

Figure 20. The velocity profiles at different downstream locations of the hump

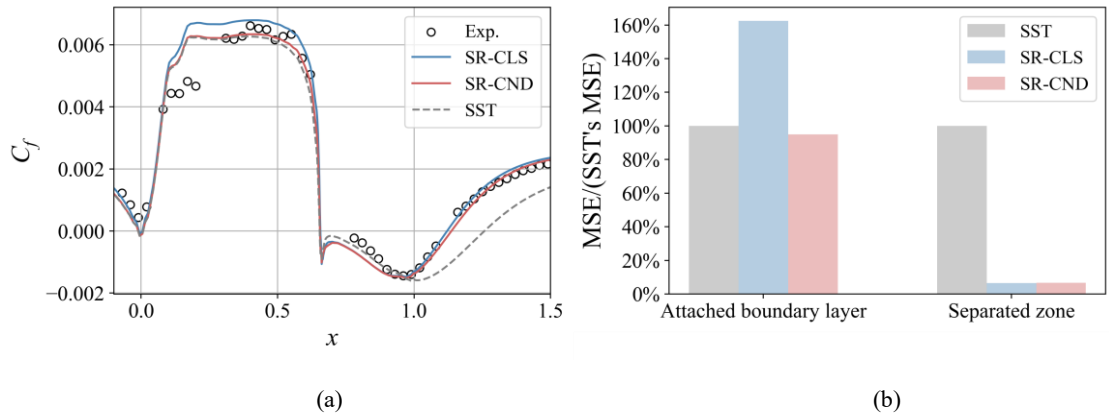

Figure 21. (a) $C_f$ distribution given by [55] (b) Relative error in each region

The $\beta$ distribution in Figure 22 shows that both the SR-CLS and the SR-CND models increase $\beta$ in the separated shear layer. Similar to its training data produced by FI-CLS, SR-CLS gives an increase in $\beta$ near the top of the hump. The increase in $\beta$ is responsible for the increased MSE of $C_f$ on the top of the hump, as shown in Figure 21(b).

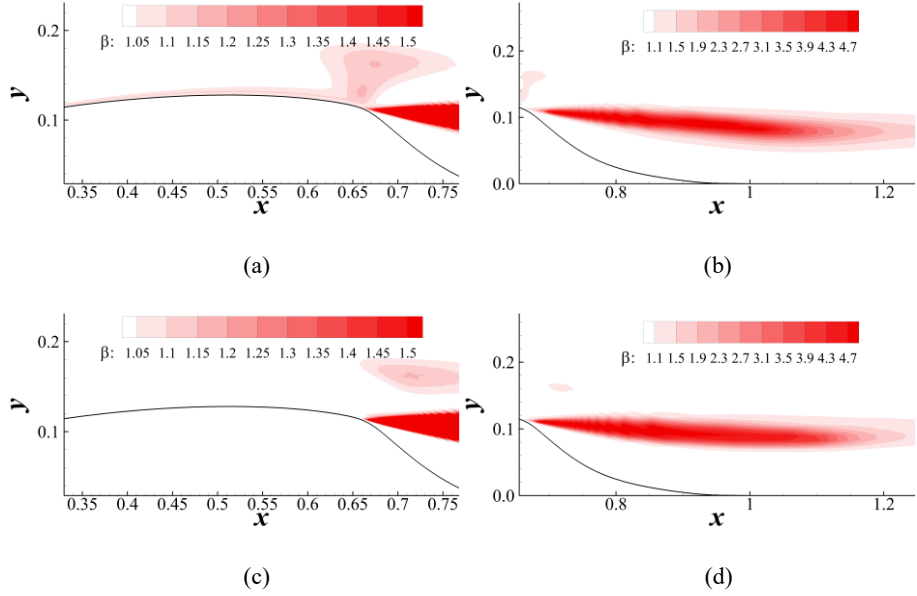

Figure 22. $\beta$ distributions: (a) and (b) are obtained via SR-CLS, and (c) and (d) are obtained via SR-CND. (a) and (c) are on the top of the hump, while (b) and (d) are downstream of the hump.

In summary, the SR-CLS and the SR-CND models can improve the baseline SST model performance in terms of separated flow to a similar level, but the SR-CLS model degrades the baseline SST model accuracy in the attached boundary layer, while the SR-CND model preserves it. This is different from the CBFS case, where the SR-CLS model and the SR-CND model provide similar results in the attached boundary layer. This difference might be caused by the different Reynolds numbers in the two cases.

## 4. Differences between test cases and the training set

This section evaluates the SR-CLS and SR-CND models across various unseen scenarios. Both models exhibit superior performance in predicting separation flow compared to the baseline SST model in most cases tested, showing similar and relatively good L3 generalizability. However, only the SR-CND model developed by conditioned field inversion maintains the accuracy of the SST model in the attached boundary layer (e.g., the ZPG flat plate, NACA0012 at zero angle of attack), demonstrating effective L2 generalizability.

## 4.1. Periodic hills

In this section, we apply the SR-CLS and SR-CND models to classic periodic hills. The geometry construction and DNS analysis were carried out by Xiao et al. [56]. $\alpha$ is a parameter that controls the aspect ratio of the periodic hills, as shown in Figure 23. The Reynolds number based on the hill height is 5600. 3 levels of grids are applied to test the grid convergence property. The height of the first grid layer near the wall ensures $\Delta y^+ < 1$. All computations are carried out using OpenFOAM's SimpleFOAM solver. The friction coefficients defined by Eq.(21) ($U_b$ is the bulk velocity, $L$ is the streamwise length of the geometry, and the integration is taken on both the upper surface and the lower surface) on different levels of grids are plotted in Figure 24. The $C_F$ difference between 3 levels of mesh are all smaller than $1 \times 10^{-4}$, with the difference between the medium mesh and the fine mesh around $5 \times 10^{-5}$. Consequently, the grid convergence property is satisfying.

$$C_F = \frac{1}{\frac{1}{2}\rho U_b^2 L} \int C_f dx \tag{21}$$

The velocity profiles given by different models are shown in Figure 25. The velocity profiles given by the same model on different grid levels almost overlap, showing good grid convergence properties. The SR-CND and the SR-CLS models both improve the baseline SST model's result in each case, with the SR-CND model giving a relatively better result in the $\alpha = 1.0$ case. Figure 26 suggests that both the SR-CND and the SR-CLS models also bring some enhancement to the accuracy of $C_f$ prediction.

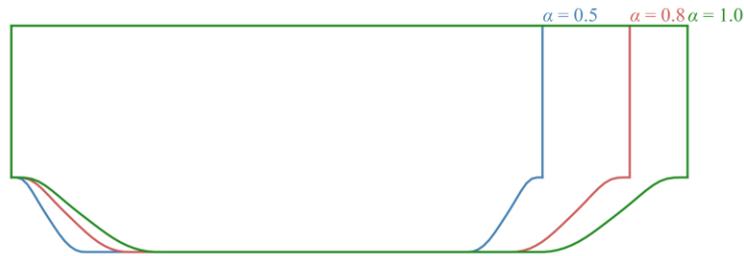

Figure 23. The geometry of the periodic hills

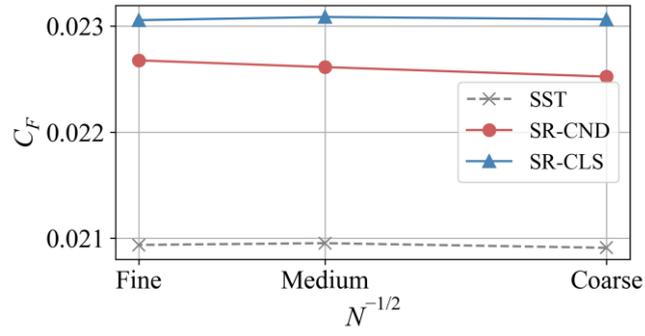

Figure 24. Friction coefficients obtained by different models on 3 levels of meshes, $\alpha = 1.0$.

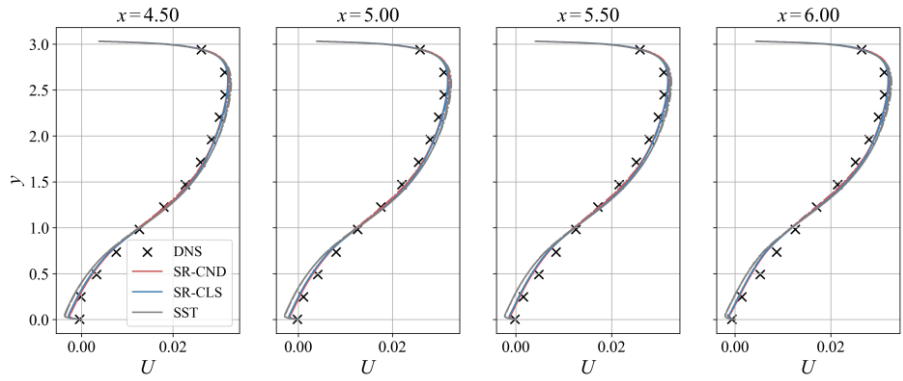

(a)

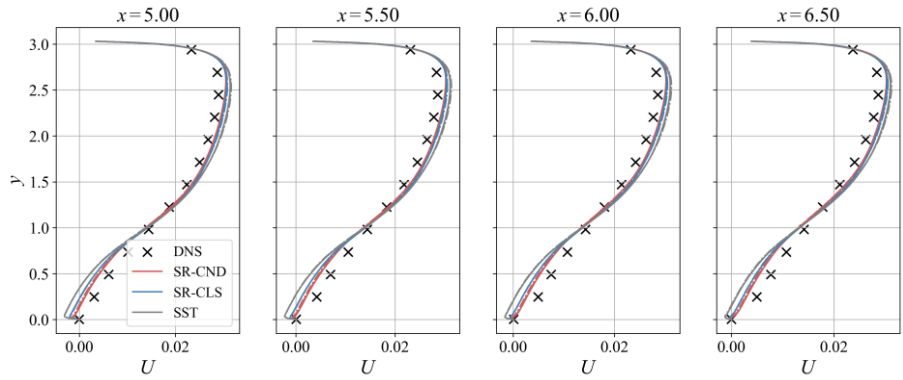

(b)

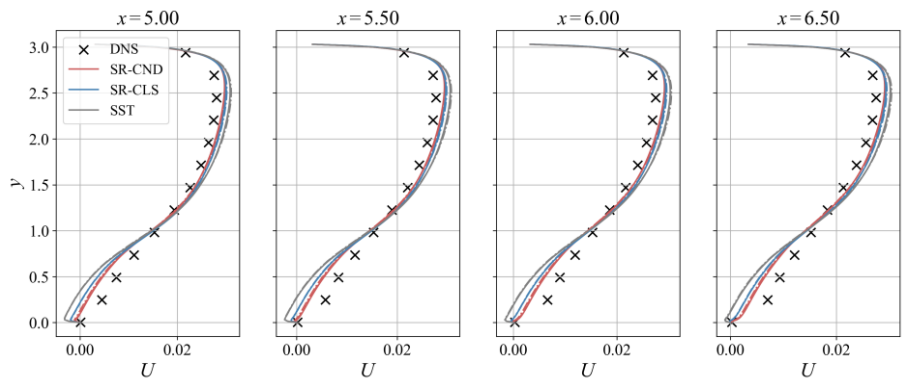

(c)

Figure 25. Velocity profiles of the (a) $\alpha = 0.5$ case, (b) $\alpha = 0.8$ case, and (c) $\alpha = 1.0$ case. The dash-dot line represents the coarse mesh, the dashed line represents the medium mesh, and the solid line represents the fine mesh.

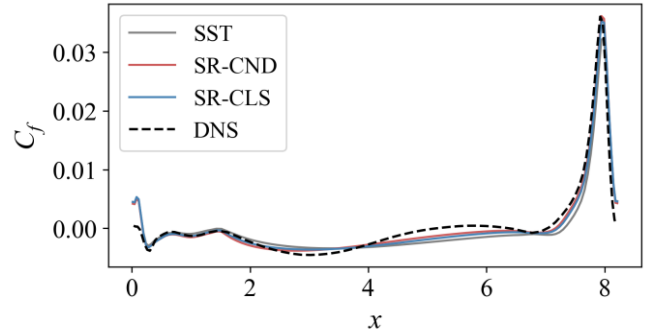

Figure 26. The friction coefficient distribution of $\alpha = 0.8$, fine mesh along the lower surface given by different models

For simplicity, we only show the streamline and $\beta$ distribution of the $\alpha = 0.8$ case on fine mesh here in Figure 27. The other cases are similar. The SST model predicts that the flow will attach to the next hill. On the other hand, the reattachment point given by the SR-CND and the SR-CLS models is approximately $x = 6.0$, which is much closer to the DNS result ($x \approx 5.2$) than the SST model. The $\beta$ given by the SR-CND model concentrates near the separated shear layer, while the $\beta$ distribution of the SR-CLS model is more diffusive. The SR-CLS activates $\beta$ near the upper surface, where the flow remains attached and might not need any correction.

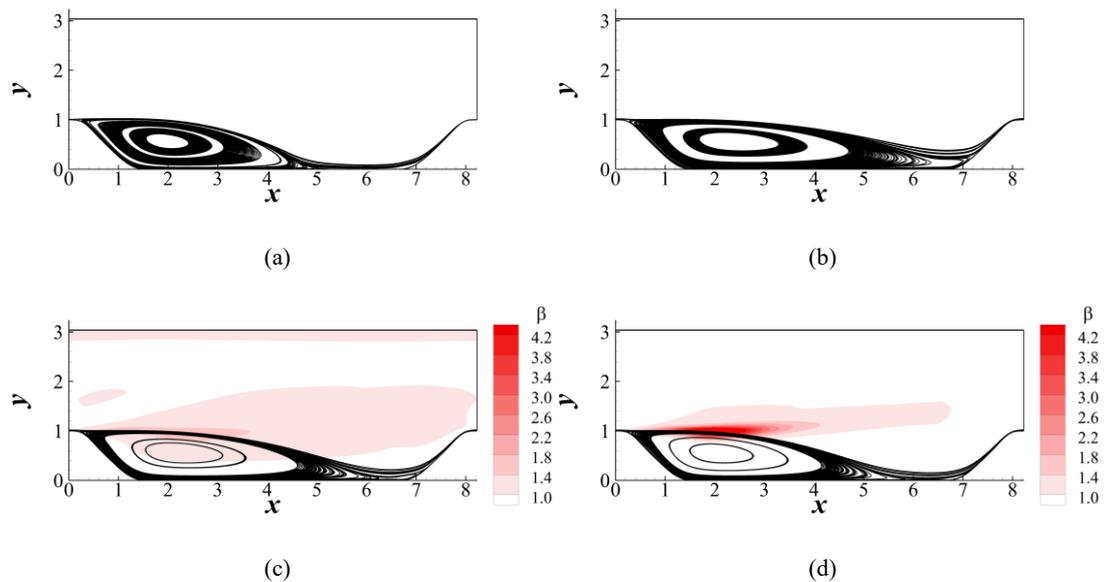

Figure 27. Streamline plot of (a) DNS, (b) SST, (c) SR-CLS, with $\beta$ contour, (d) SR-CND, with $\beta$ contour

In summary, the SR-CND model and the SR-CLS model have similar abilities to predict separated flow in a series of periodic hill cases. Both of the models substantially outperform the baseline SST model.

## 4.2. The NLR-7301 two-element airfoil

The prediction of aerodynamic performance, particularly stall behavior, in multiple-element airfoils is crucial for designing high-lift configurations. The SST model often inaccurately predicts a large separation zone near the wake [57], which is a key issue in stall behavior resolution. In this section, we assess whether the SR-CLS and SR-CND models can enhance stall prediction compared to the baseline SST model using the NLR-7301 two-element airfoil [59][60] as a test case. Figure 28 illustrates the geometry of the NLR-7301 airfoil and its C-type mesh. The setup involves a Reynolds number of $2.51 \times 10^6$ based on the chord length and a freestream Mach number of 0.185. The height of the first grid layer is adjusted to ensure that $\Delta y^+$ remains below 1. Given the potential for the Mach number near the airfoil's leading edge to reach 0.6 at high angles of attack, we employ the compressible flow solver CFL3D [58] for this analysis.

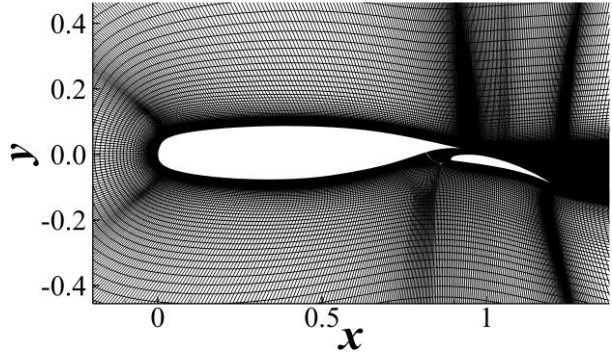

Figure 28. The computational mesh of the NLR-7301 airfoil.

Figure 29(a) displays the lift coefficient versus the angle of attack ($C_L - AOA$) curves generated by various models alongside experimental data [60]. As shown in Table 5, both the SR-CLS and SR-CND models produce results more aligned with the experimental values, with the SR-CND model achieving a maximum lift coefficient ($C_{L,max}$) error of just 1%. In comparison, the SR-CLS model exhibits a larger relative error in $C_{L,max}$, and the baseline SST model underestimates both $C_{L,max}$ and the stall angle of attack ($AOA_{stall}$). Figure 29(b) presents the pressure coefficient ($C_p$) distribution at $AOA = 13.1°$. Here, the SST model predicts a lower $C_p$ on the suction surface than do the other models. While both the SR-CLS and SR-CND models correlate well with the experimental data, the SR-CLS model shows a slightly more pronounced underestimation of the suction peak. Figure 29(c) shows the $C_f$ distribution at $AOA = 13.1°$. The data on the flap is shifted by 0.1 downstream for clarity. The result obtained by the SST model shows a clear separation near the trailing edge while the SR-CLS and the SR-CND model predict no separation. The comparison with the experiment is rather qualitative here because the experiment $C_f$ data is very sparse. In Figure 30, the Mach number contours and streamlines are illustrated. The baseline SST model predicts a significant separation zone starting from the rear part of the main wing. Conversely, the SR-CLS and SR-CND models show no such flow separation.

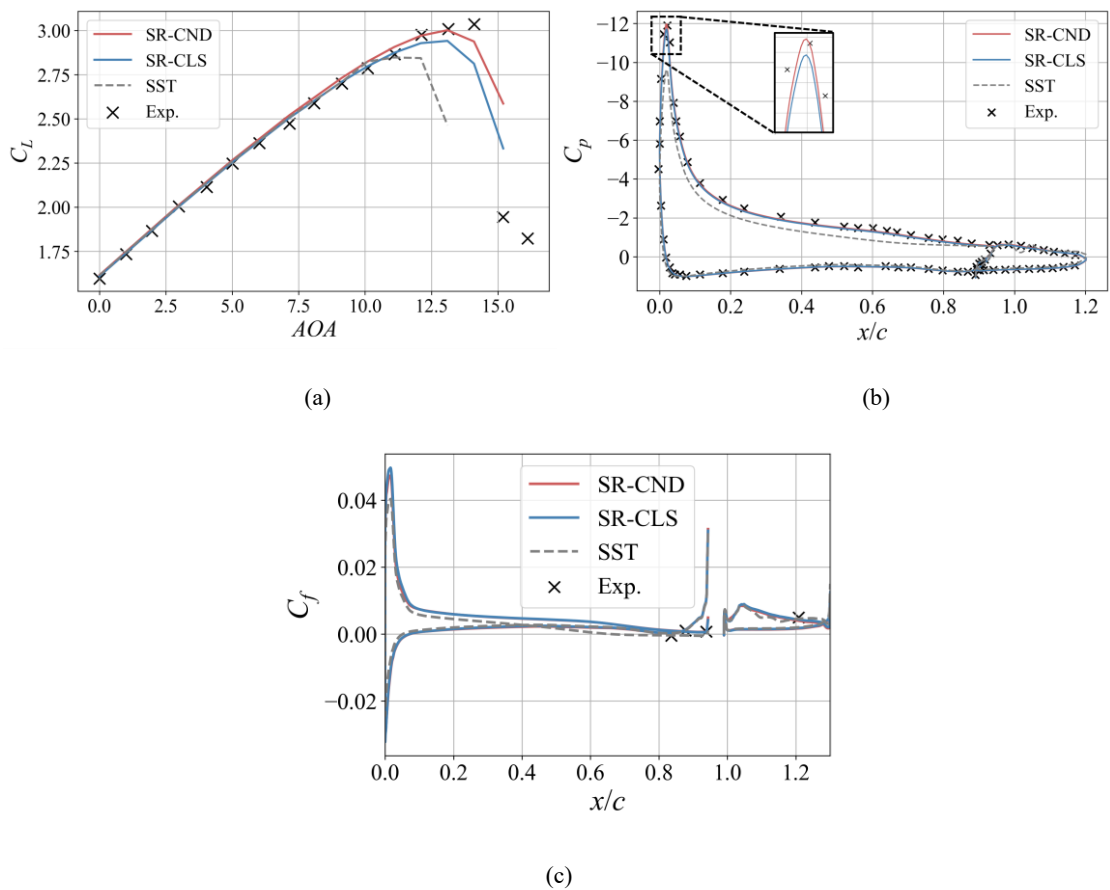

Figure 29. The aerodynamic performance of NLR-7301 given by the SST, SR-CLS, and SR-CND models: (a) $C_L - AOA$ curve, (b) $C_p$ distribution at $AOA = 13.1°$, (c) $C_f$ distribution at $AOA = 13.1°$

Table 5. The results and relative errors of different models

|  | Experiment [60] | SST | SR-CLS | SR-CND |
|---|---|---|---|---|
| $C_{L,max}$/relative error | 3.03/-- | 2.84/6.3% | 2.94/3% | 3.00/1% |
| Stall $AOA$/deviation | 14.1°/-- | 11.1°/3° | 13.1°/1° | 13.1°/1° |

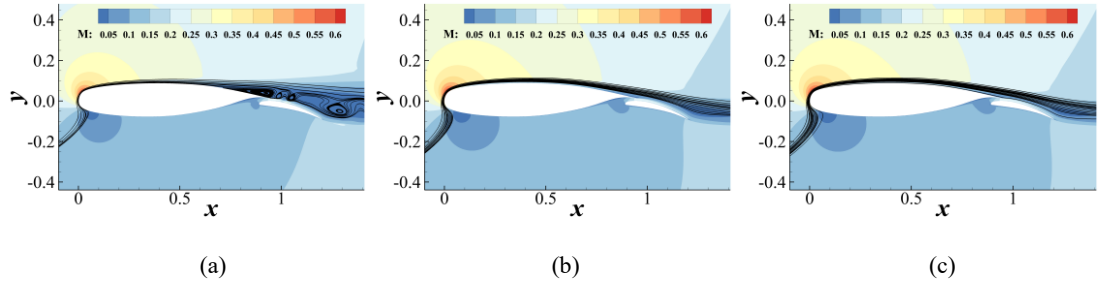

Figure 30. The Mach number contour and the streamlines at $AOA = 13.1°$ given by (a) the SST model, (b) the SR-CLS model, and (c) the SR-CND model.

Figure 31 indicates that for both the SR-CLS and SR-CND models, $\beta$ increases in two key mixing regions, as identified in Figure 32. The first mixing region is where the free-shear layer, originating from the leading edge of the main wing at high $AOA$, meets the wake. The second mixing region is created by the interaction of the flap's wake with the jet from the slot. We analyze the physical mechanisms behind the suppressed separation observed in Figure 30 for both models. In these mixing regions, $\beta$ is elevated due to strong shear forces, leading to increased $|\lambda_2|$ and $|\lambda_5|$ values, as indicated in Table 2 and Table 4. This increase in $\beta$ amplifies the destruction of $\omega$ in these regions, resulting in reduced dissipation of $k$ and, consequently, a higher turbulent viscosity ($\nu_T$). This enhanced $\nu_T$, compared to that predicted by the baseline SST model, facilitates more effective momentum diffusion from the mainstream to the shear layer. As a result, the shear layers modeled by SR-CLS and SR-CND possess greater momentum, making them less prone to flow separation than those predicted by the baseline SST model.

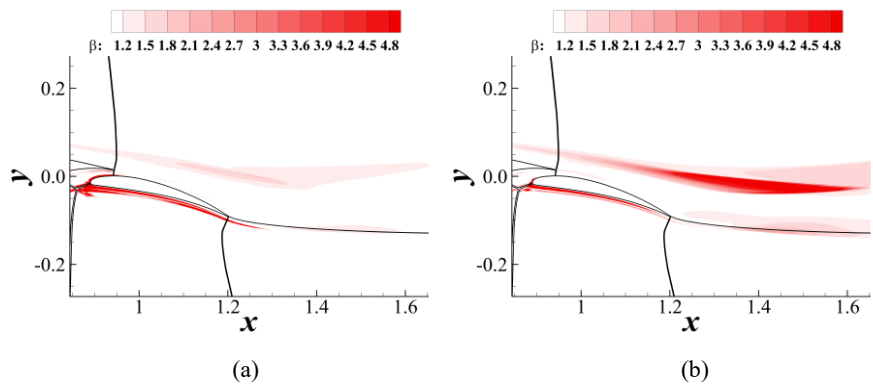

Figure 31. $\beta$ distribution at $AOA = 13.1°$ given by (a) the SR-CLS model and (b) the SR-CND model

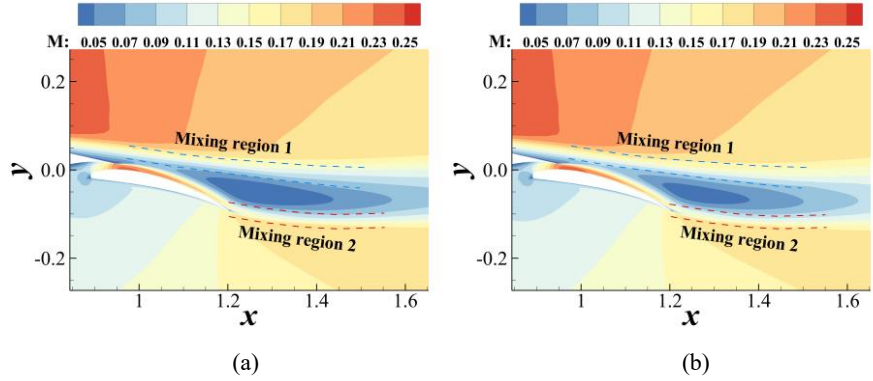

(a) (b)

Figure 32. The structure of the flow field near the flap at $AOA = 13.1°$ given by (a) the SR-CLS model and (b) the SR-CND model

In summary, both the SR-CLS and SR-CND models outperform the baseline SST model in predicting stall behavior for the NLR-7301 airfoil, a scenario distinct from their training set, demonstrating good L3 generalizability in this test case. Notably, the SR-CND model achieves an impressively low relative error of just 1% in $C_{L,max}$ compared to the experimental data. Furthermore, these models maintain stability and effectiveness when integrated with the compressible CFD code CFL3D, highlighting their strong portability and applicability in various aerodynamic analyses.

## 4.3. SAE notchback standard model

In this part, we apply the SR-CLS and the SR-CND model to a simplified notchback car model to test their ability to generalize to 3-D complex separated flow. The geometric model used here is often referred to as the SAE model [62]. The experimental data used in this part were obtained from [62]. The computational domain is shown in Figure 33(a), which resembles the experimental environment in [62]. $L$ in Figure 33(a) refers to the length of the SAE model. The inlet velocity is $40\ m/s$, and the Reynolds number based on the model length is $2.3 \times 10^6$; both of these settings agree with the experiment. We use an unstructured mesh for this case. The grid convergence property is studied in this case. A coarse mesh with about $1 \times 10^7$ cells, a medium-mesh with about $1.8 \times 10^7$ cells, and a fine mesh with about $3.3 \times 10^7$ cells are constructed. All meshes are built upon the half model. The grid point distributions for the coarse mesh on the symmetry plane and the SAE body are shown in Figure 33(b) and (c). More cells are added to the notchback to capture the potential tiny separation there. Computations are performed using OpenFOAM's

SimpleFOAM solver. Figure 34 shows that for the SR-CND model and the SR-CLS model, the drag coefficients calculated on the fine and the medium are essentially the same (<0.001). For the SST model, the drag difference between the fine mesh and the medium mesh is slightly larger (~0.001) but is also substantially smaller than the drag difference between the coarse mesh and the medium mesh (~0.005). Consequently, the grid convergence is satisfactory. In the subsequent discussion of this section, we plot the result of the medium mesh.

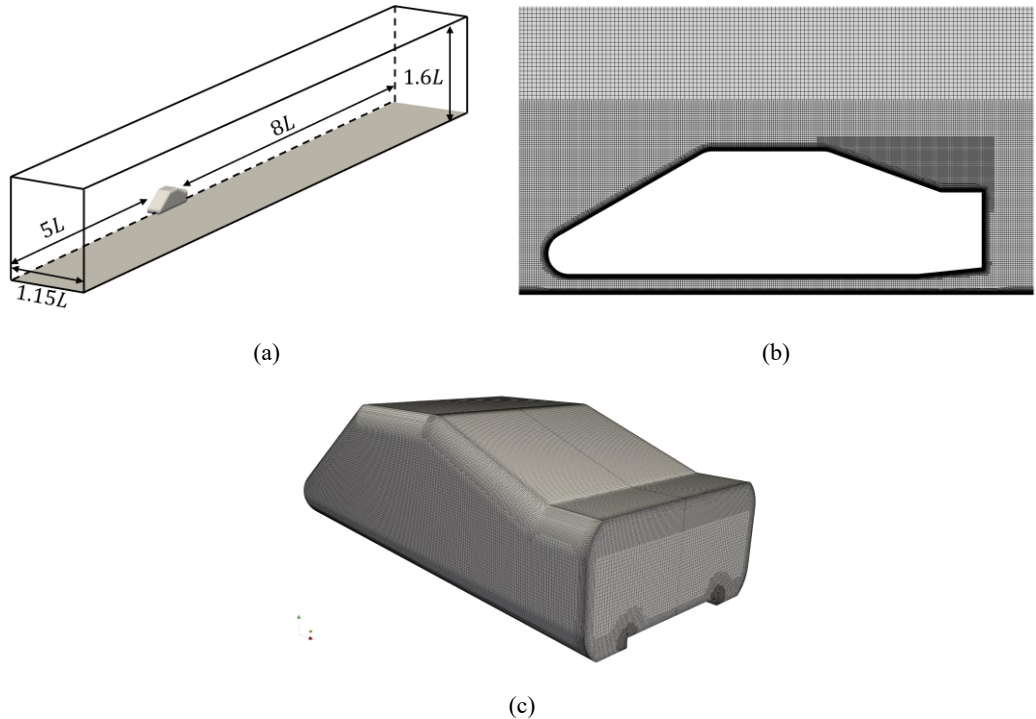

Figure 33. (a) The computational domain; (b) the grid in the symmetry plane; (c) the surface mesh of the model

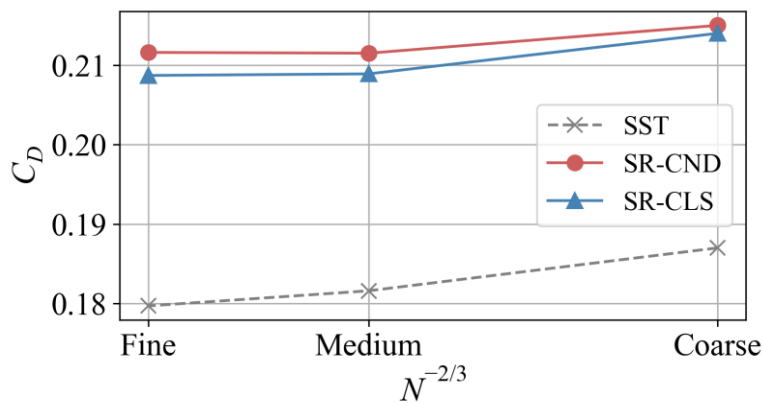

Figure 34. The grid convergence properties of different models.

The streamlines at the rear are shown in Figure 35. The separation zones obtained by the SR-CLS, SR-CND, and PIV experiments all ended at approximately $x = 0.60\ m$. On the other hand, the SST model overpredicts the recirculation, giving a separation zone ending at $x \approx 0.63\ m$. The velocity profiles shown in Figure 36 also indicate that the SR-CLS and SR-CND models outperform the baseline SST model and align better with the PIV data. The velocity profiles obtained on the coarse mesh, the medium mesh, and the fine mesh have relatively small differences, indicating good grid convergence. Figure 37 shows that both the SR-CLS and the SR-CND models increase $\beta$ in the separated shear layers, thus suppressing the separation size. Table 6 suggests that both the SR-CLS and the SR-CND models substantially decrease the error in the predicted drag coefficient.

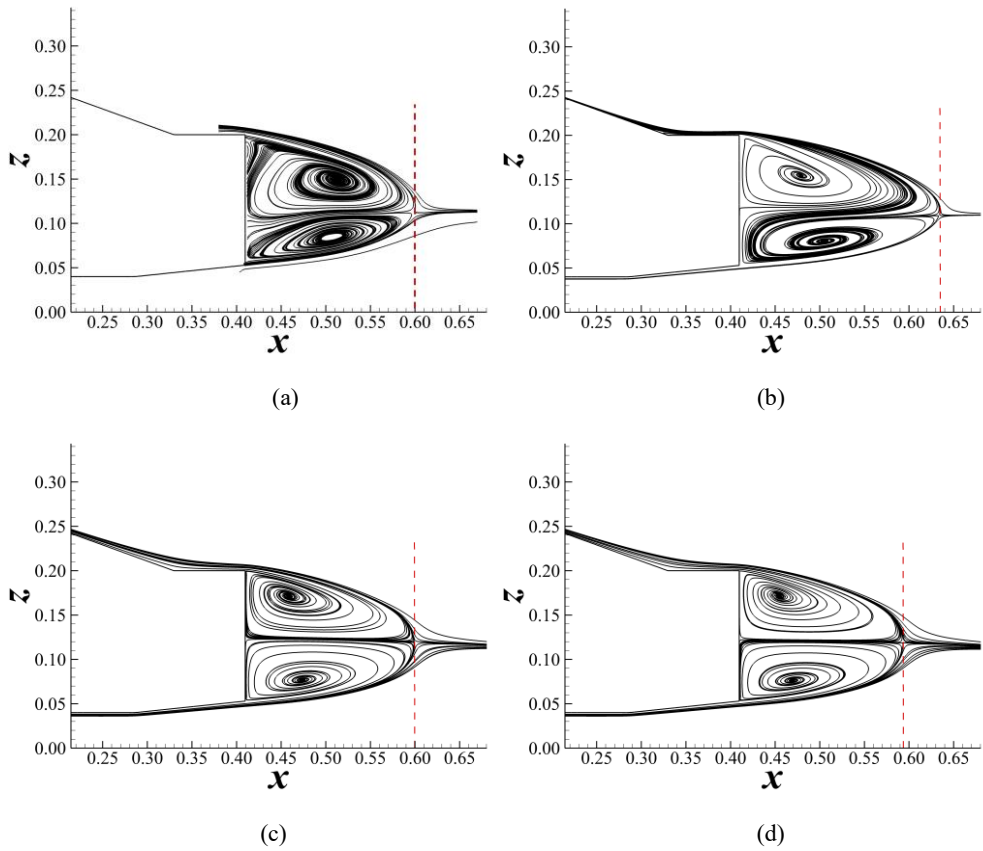

Figure 35. Separation bubble at the rear of the car obtained by (a) PIV[62], (b) SST, (c) SR-CLS, and (d) SR-CND. The results on the medium mesh are shown.

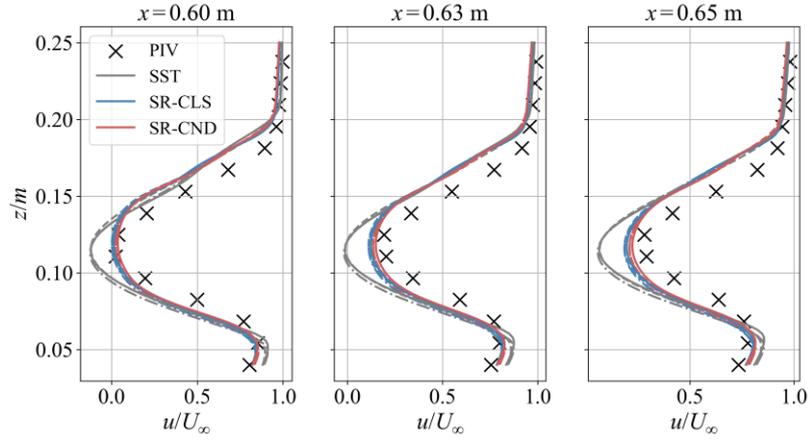

Figure 36. The velocity profiles near the end of the separation zone at the rear obtained on different grids. The dashed line indicates the coarse mesh, the dash-dot line indicates the medium mesh and the solid line indicates the fine mesh.

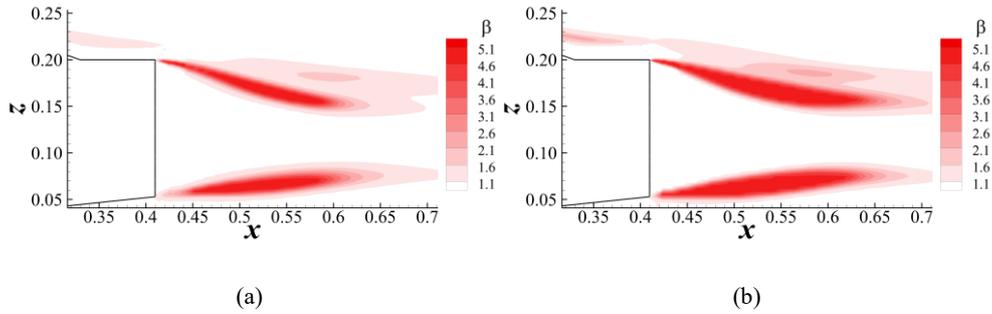

(a)          (b)

Figure 37. $\beta$ distribution in the wake obtained by (a) SR-CLS and (b) SR-CND. The results on the medium mesh are shown.

Table 6. Drag coefficients ($C_D$) obtained by different models (medium mesh)

|  | **SST** | **SR-CND** | **SR-CLS** | **Experiment** |
|---|---|---|---|---|
| $C_D$ | 0.181 | 0.211 | 0.209 | 0.207 |
| Relative error | 12.6% | 1.9% | 1.0% | -- |

Figure 38 illustrates the $\beta$ distribution near the end of the notchback. Here, the SR-CLS model shows an increase in $\beta$ close to the wall, whereas the $\beta$ value in the SR-CND model remains at 1.0 near the wall, a result of incorporating the $f_d$ function. In Figure 39, the separation region (marked by the $u_x = -0.001 \, m/s$ contour) reveals that both the SR-CLS and SR-CND models reduce the thin separation zone at the notchback end. The SR-CLS model is more effective at suppressing this separation due to its increased $\beta$ near the wall. However, as shown in Figure 40, this increased $\beta$ near the wall in the SR-CLS model leads to significant deviations in the pressure coefficient ($C_p$) distribution on the upper surface of the SAE model, particularly near the notchback end, diverging notably from both the SST model and experimental data. This deviation is caused by the reduced thin separation predicted by the SR-CLS model. In contrast, the SR-CND model also introduces some errors in $C_p$, these errors are relatively minor compared to those of the SR-CLS model. This more accurate performance of the SR-CND model is attributed to its mechanism for protecting the near-wall region.

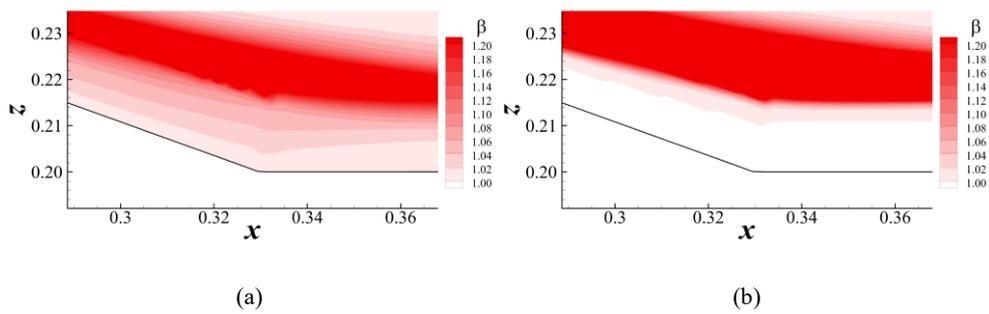

Figure 38. $\beta$ distribution at the end of the notchback given by the (a) SR-CLS and (b) SR-CND methods. The results on the medium mesh are shown.

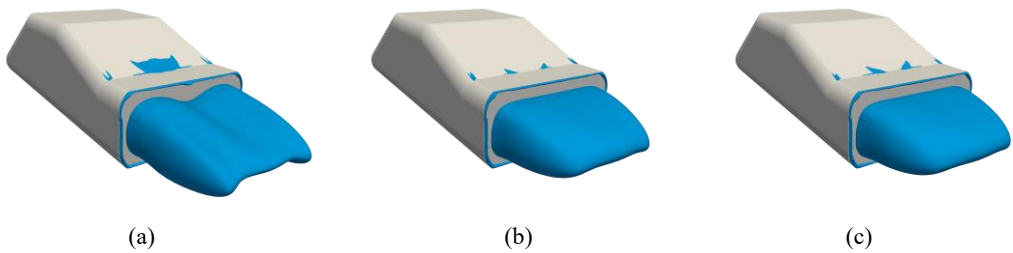

Figure 39. Contours of $u_x = -0.001 \, m/s$ given by (a) SST, (b) SR-CLS, and (c) SR-CND. The results on the medium mesh are shown.

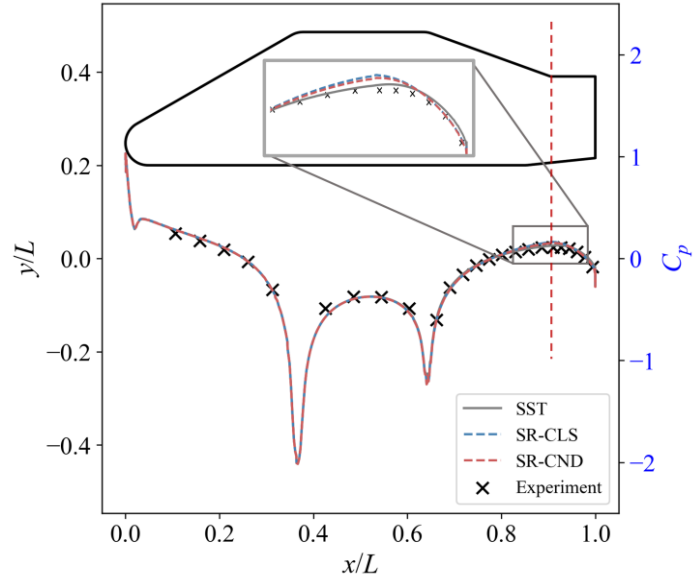

Figure 40. $C_p$ distribution on the upper surface given by different models and the experiment. The results on the medium mesh are shown.

In summary, the SR-CLS and the SR-CND models can both improve the accuracy of the baseline SST model for 3-D separated flows around the SAE, showing good generalizability to L3. However, the negative impact of the SR-CND model on the region where the baseline model already performs well, i.e., the end of the notchback, is smaller than that of the SR-CLS model.

## 4.4. Ahmed body standard model

In this section, we use the SR-CLS model and the SR-CND model to compute the 3-D separated flow around an Ahmed body with a slant angle of 25.0 degrees [63]. In contrast to the SAE model in the previous section, which resembles a sedan, the geometry of the Ahmed body's rear is more similar to that of a sport utility vehicle (SUV) or a hatchback vehicle. The half model shown in Figure 41(a) is analyzed in this study. The Reynolds number based on the body length is $2.78 \times 10^6$, and the inlet velocity is $40\ m/s$. Different from the previous 3D case, we use structured mesh here, as shown in Figure 41(b). Three meshes are used to test the grid convergence property. The coarsest mesh has approximately $3.6 \times 10^6$ cells, the medium mesh has around $7.2 \times 10^6$ cells and the fine mesh has about $1.4 \times 10^7$ cells. The CFD analysis in this section is conducted using OpenFOAM's SimpleFOAM solver.

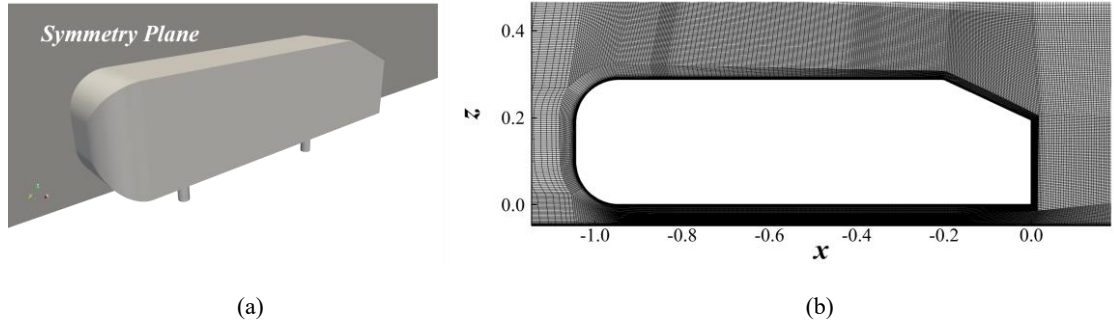

(a)                                            (b)

Figure 41. The configuration of the Ahmed body case. (a) The half model and the symmetry plane. (b) The medium computational mesh

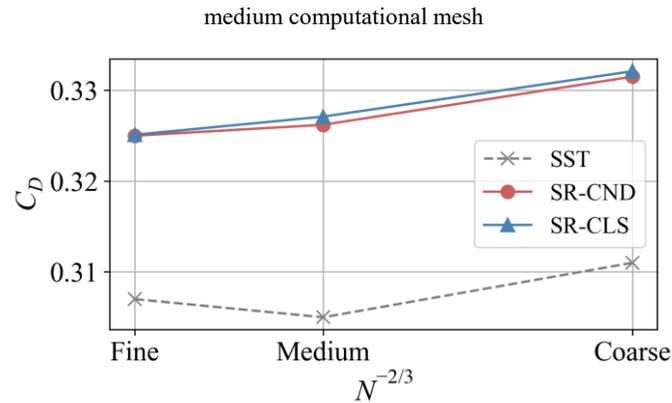

Figure 42. The drag coefficients obtained by different models on 3 levels of mesh

For the SR-CND and the SR-CLS model, the difference in drag coefficient between the coarse mesh and the medium mesh is about 0.005, and the difference between the medium mesh and the fine mesh is 0.001~0.002, demonstrating good grid convergence properties. The drag coefficient fluctuates as the mesh becomes denser for the SST model. Figure 43 shows that the SR-CND model and the SR-CLS model give more accurate predictions for the velocity profiles than the baseline SST model when compared to the PIV data [64]. The velocity profiles given by the same model on different grids essentially overlap, showing good grid convergence properties. The mean square error between the predicted velocity (normalized by the freestream velocity $U_\infty = 40\ m/s$) and the PIV data [64] is shown in Table 7. Both the SR-CND and the SR-CLS models reduce the MSE by approximately 90%. Figure 44 (a) and (b) show that the SR-CND and SR-CLS models increase $\beta$ near the separated shear layers originating from the upper surface and the lower surface. The pattern of $\beta$'s distribution is similar to that of the SAE case. As shown in Figure 44 (c) and (d), the SR-CND model and the SR-CLS model both increase $\beta$ slightly near the slant, but only the SR-CLS model activates $\beta$ immediately near the wall. The 3-D streamline plots in Figure 45 show similar separation patterns given by the SR-CLS and the SR-CND models, where the flow remains attached all along the slant, agreeing with the experiment [64]. However, the separation structure given by the SST model is completely different, with a large separation starting from the beginning of the slant.

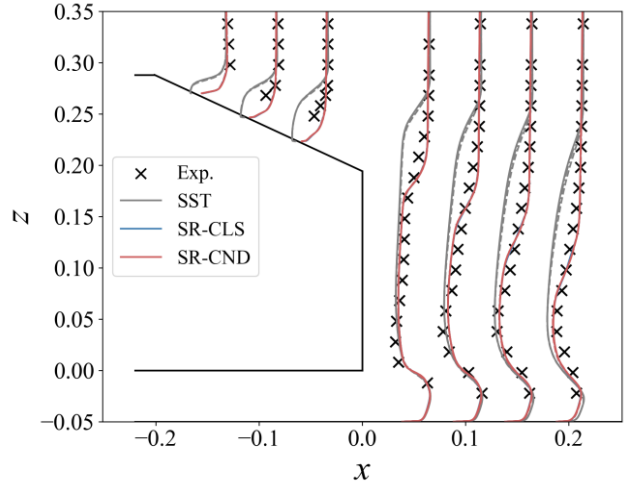

Figure 43. Velocity profiles given by different turbulence models on different grid levels. The dashed line represents the coarse mesh, the dash-dot line represents the medium mesh, and the solid line represents the fine mesh.

Table 7. The MSEs of the velocity profiles given by different turbulence models calculated on the fine mesh

|  | *SST* | *SR-CND* | *SR-CLS* |
| --- | --- | --- | --- |
| Velocity profile's RMSE | 0.677 | 0.072 | 0.069 |
| Relative to baseline (SST) | 100.0% | 10.6% | 10.2% |

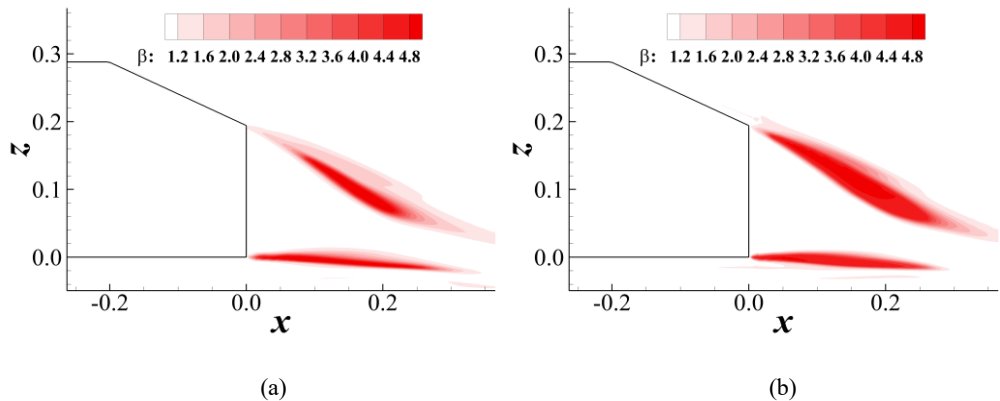

(a)  (b)

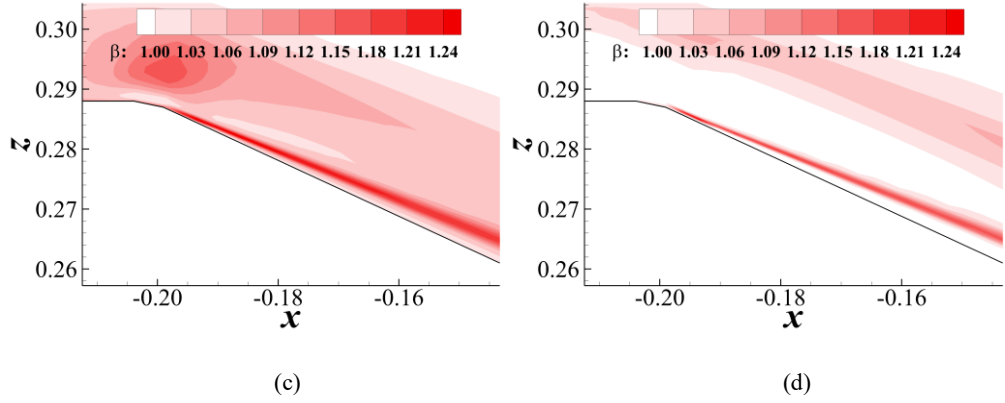

Figure 44. $\beta$ distributions (a) given by the SST-CLS model in the wake, (b) given by the SR-CND model in the wake, (c) given by the SST-CLS model near the slant, and (d) given by the SR-CND model near the slant. The results on the fine mesh are shown.

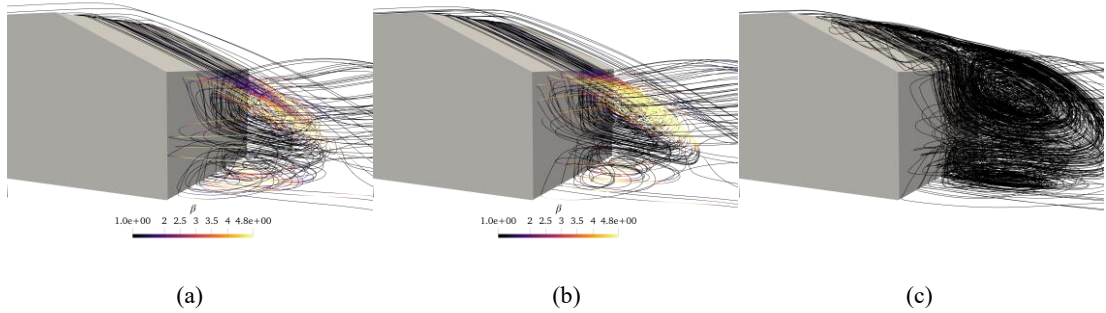

Figure 45. Streamlines in the separation zone given by (a) SR-CLS, (b) SR-CND, and (c) SST. The streamlines given by the SR-CLS and the SR-CND model are colored based on $\beta$. The results on the fine mesh are shown.

In conclusion, the SR-CLS and the SR-CND models successfully generalize to this 3-D SUV-like case, outperforming the baseline SST model in terms of separation prediction.

## 4.5. Attached boundary layer tests

Our primary goal in introducing conditioned field inversion was to maintain the original calibration of the model in wall-attached flows. Earlier tests have shown that the SR-CND model performs comparably to the SR-CLS model in separated flows, demonstrating similar L3 generalizability. Notably, the SR-CND model does not enable the corrective factor $\beta$ near the wall in these cases. To assess whether our initial objective is achieved, we directly evaluate both the SR-CLS and SR-CND models on attached boundary layers. This evaluation includes tests on both the zero-pressure-gradient (ZPG) flat plate boundary layer and the pressure-gradient boundary layer on the

NACA0012 airfoil using the SimpleFOAM solver in OpenFOAM.

We first introduce the test on the zero-pressure-gradient boundary layer. The computational domain and the boundary conditions are shown in Figure 46. The Reynolds number based on the flat plate length is $Re_L = 1.0 \times 10^7$. A grid containing approximately $2 \times 10^5$ rectangular cells are used. The height of the first grid layer ensures that $\Delta y^+ \leq 0.05$. Figure 47(a) shows that the friction coefficients given by the baseline SST model and the SR-CND model agree well with the experimental results [66], while the SR-CLS model overestimates the skin friction. Figure 47(b) demonstrates that the baseline SST model and the SR-CND model capture the viscous layer and the log layer with good accuracy, but the SR-CLS model yields an erroneous log layer. Figure 47(b) also suggests that the boundary layer thickness is overpredicted by the SR-CLS model compared with the SST result.

Figure 48 shows the $\beta$ distributions given by the SR-CLS model and the SR-CND model at $x = 0.3$. The SR-CND gives a uniform $\beta$ distribution of 1, while the SR-CLS model increases $\beta$ to approximately 1.06. Although the increase in $\beta$ is relatively small, the error caused by it in the boundary layer in terms of the friction and velocity profile is not negligible.

In conclusion, the model generated by the classic field inversion method (the SR-CLS model) indeed breaks the calibration by increasing $\beta$ in the near-wall region and therefore puts the accuracy of the ZPG boundary layer at risk. On the other hand, the model generated by the proposed method, namely, the SR-CND model, can maintain $\beta = 1$ in the ZPG boundary layer, thus preserving the calibration and accuracy of the baseline SST model.

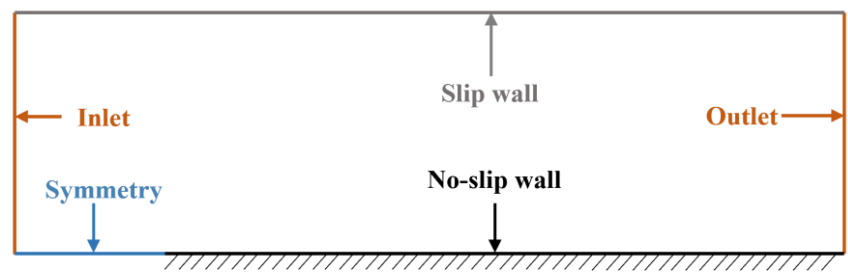

Figure 46. The computational domain of the ZPG flat plate

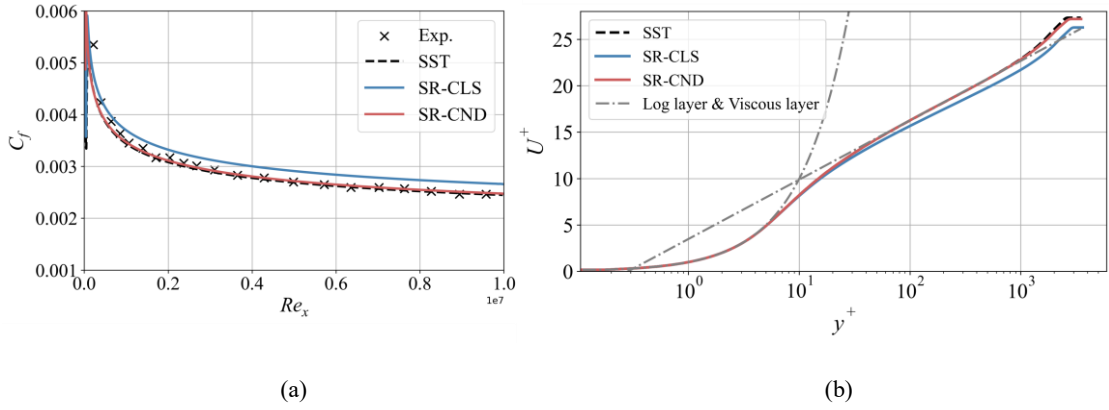

(a)  (b)

Figure 47. (a) Friction coefficient $C_f$ distribution and (b) velocity profiles at $Re_x = 0.5 \times 10^7$

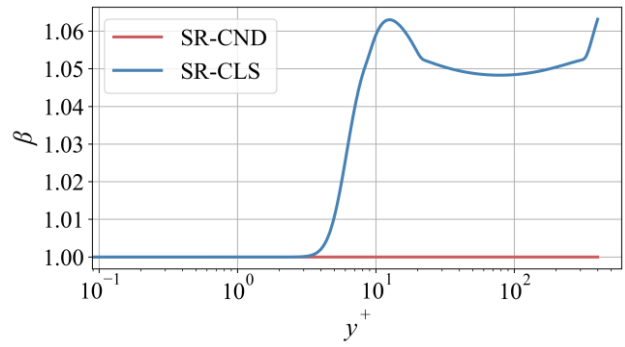

Figure 48. $\beta$ distribution along the wall-normal direction at $Re_x = 1.5 \times 10^6$

Now, we introduce the test on the NACA0012 airfoil at a zero angle of attack, in which the baseline SST model is thought to be sufficiently accurate for skin friction prediction [65]. Due to the curvature of the airfoil surface, a pressure gradient is exerted on the attached boundary layer. The Reynolds number based on the chord length is $Re_c = 6 \times 10^6$, and the Mach number is approximately 0.15. We use the C-type grid shown in Figure 45 and ensure that $\Delta y^+ \approx 0.6$ for the first grid layer near the wall. A total of 640 cells are distributed along the airfoil surface, and 120 cells are deployed in the wake. A total of 136 cells are used along the wall-normal direction. The $C_f$ distribution is shown in Figure 50. Compared with the baseline SST model, the SR-CND model gives nearly identical results, while the SR-CLS model overpredicts skin friction.

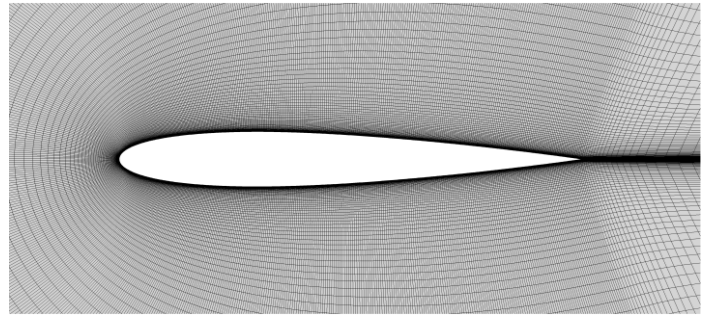

Figure 49. The C-type mesh used for NACA0012

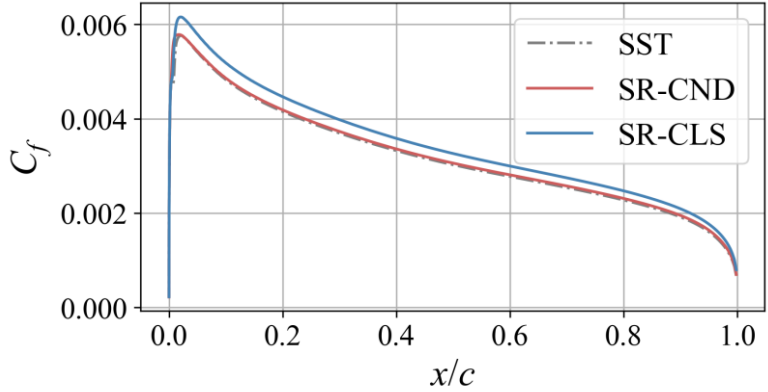

Figure 50. $C_f$ distribution along the airfoil surface

## 5. Conclusions

In our research, we introduced an innovative approach known as conditioned field inversion, which integrates a shielding function, $f_d$, into the corrective factor $\beta$ in the $\omega$ equation. This addition is designed to maintain the original calibration of the model in the boundary layer during the field inversion process. We applied both the conditioned and classic field inversion methods to the NASA hump and the CBFS case. Two models were then developed: the SR-CLS model based on the classic field inversion dataset and the SR-CND model derived from the conditioned field inversion dataset. We then evaluated these models across various test scenarios, including separated and wall-attached flows. Our findings lead to several key conclusions:

1. Both the SR-CLS and SR-CND models demonstrated relatively good performance in separated flows presented in the paper. The results show that the conditioned field inversion method does not compromise the ability of the data-driven models to generalize to different separated flows with similar physical mechanisms (L3 generalizability).

2. The SR-CND model, developed using the conditioned field inversion method, matched the accuracy of the baseline SST model in scenarios where the baseline model was already effective, such as the ZPG flat plate and the NACA0012 airfoil at a zero angle of attack. This signifies a high level of L2 generalizability. Conversely, the SR-CLS model, trained with the classic field inversion method, failed to maintain this calibration and showed significant errors, indicating poor L2 generalizability.

In summary, our conditioned field inversion method enables the creation of models that not only have similar L3 generalizability compared with the models created by the classic field inversion method but also preserve the accuracy of the baseline model for wall-attached flows (L2 generalizability). As demonstrated in our study, achieving this dual capability is challenging with the classic field inversion method. This highlights the advancements our method has made in the field of data-driven turbulence modeling.

## Acknowledgment

This work was supported by the National Natural Science Foundation of China (grant nos. 12388101, 12372288, U23A2069, and 92152301).

## Declaration of Interests

The authors report no conflict of interest.